\documentclass[11pt]{article}

\usepackage{amsmath,amssymb}

\newcommand{\R}{\mathbb{R}}
\newcommand{\C}{\mathbb{C}}
\newcommand{\G}{{\cal G}}

\newcommand{\pr}{{\rm pr}}
\newcommand{\g}{\gamma}
\renewcommand{\a}{\alpha}
\newcommand{\A}{{\cal A}}

\newcommand{\D}{{\cal D}}
\newcommand{\Abar}{\overline{\cal A}}

\renewcommand{\S}{{\cal S}}
\newcommand{\Z}{\mathbb{Z}}
\newcommand{\N}{\mathbb{N}}
\newcommand{\h}{{\cal H}}
\newcommand{\scal}[2]{\langle #1| #2\rangle}
\newcommand{\bra}[1]{\langle #1|}
\newcommand{\ket}[1]{| #1\rangle}
\newcommand{\im}{{\rm im\,}}

\newcommand{\cyl}{{\rm Cyl}}
\newcommand{\Cyl}{{\rm Cyl}}

\newcommand{\sca}{\mathfrak{A}}

\newcommand{\id}{{\rm id}}

\newcommand{\ot}{\otimes}

\newcommand{\bld}[1]{\boldsymbol{#1}}
\newcommand{\tl}[1]{\tilde{#1}}
\newcommand{\tll}{\tilde{l}}

\newcommand{\tL}{\tilde{\cal L}}
\newcommand{\la}{\lambda}
\DeclareMathOperator{\tr}{{\rm tr}}
\newcounter{mnotecount}[section]

\newtheorem{thr}{Theorem}
\newtheorem{lm}[thr]{Lemma}
\newtheorem{df}[thr]{Definition}
\newtheorem{cor}[thr]{Corollary}

\numberwithin{equation}{section}
\numberwithin{thr}{section}

\begin{document}

\title{Quantization of diffeomorphism invariant theories of connections with a non-compact structure group---an example\footnote{This is an author-created, un-copyedited version of an article accepted for publication in Communication in Mathematical Physics. The final publication is available at\newline \centerline{http://link.springer.com/article/10.1007/s00220-009-0755-y.}}}
\author{ Andrzej Oko{\l}\'ow}
\date{September 15, 2009}
\maketitle
\begin{center}
{\it  1. Institute of Theoretical Physics, Warsaw University\\ ul. Ho\.{z}a 69, 00-681 Warsaw, Poland\smallskip\\
2. Department of Physics and Astronomy, Louisiana State University,\\
Baton Rouge, LA 70803, USA\smallskip\\
oko@fuw.edu.pl}
\end{center}
\medskip

\begin{abstract}
A simple diffeomorphism invariant theory of connections with the non-compact structure group $\R$ of real numbers is quantized. The theory is  defined on a four-dimensional 'space-time' by an action resembling closely the self-dual Pleba\'nski action for general relativity. The space of quantum states is constructed by means of projective techniques by Kijowski \cite{kpt}. Except this point the applied quantization procedure is based on Loop Quantum Gravity methods. 
\end{abstract}

\section{Introduction}

General relativity (GR) expressed in the complex Ashtekar variables \cite{a-var} is a diffeomorphism invariant\footnote{A diffeomorphism invariant theory of connections means  a theory in a Hamiltonian form such that $(i)$ its configuration space is a space of connections on a principal bundle $P(\Sigma,G)$, where $\Sigma$ is a base manifold, and $G$ is a Lie group (i.e. the structure group of the bundle and the connections), $(ii)$ there exist Gauss and vector constraints imposed on the phase space which ensures, respectively, gauge and diffeomorphism invariance of the theory \cite{cq-diff}.} (background independent) theory of connections with $SL(2,\C)$ as the structure group. Canonical quantization of GR based on these variables faces two serious obstacles related to the fact that $SL(2,\C)$ is {\em non-compact} and {\em complex}:
\begin{enumerate}
\item  the non-compactness of the group makes the task of constructing the {\em space of quantum states} for the theory very difficult---by now to the best of our knowledge there is no satisfactory solution to this problem;
\item the fact that $SL(2,\C)$ is complex implies the existence of some complicated constraints called {\em reality conditions} \cite{ashtekar} which have to be included in the structure of the resulting quantum theory but, again, by now it is not clear how this should be done.
\end{enumerate}

The only solution to these problems known nowadays consists in removing the problems by removing its source, that is, the group $SL(2,\C)$: one formulates GR in terms of the Ashtekar-Barbero connection \cite{barb} whose structure group $SU(2)$ is {\em compact}, thereby the space of quantum states can be easily constructed. Moreover, since the Lie algebra of $SU(2)$ is {\em real} there are no reality conditions in this formulation. 

The above solution, however, cannot be considered as completely blameless. There are at least three reasons for that (see also \cite{sam}): $(i)$ the resulting quantum model i.e. Loop Quantum Gravity (LQG) lacks the Lorentz symmetry which was broken in a non-natural way while passing from the $SL(2,\C)$-connection to the $SU(2)$ one; $(ii)$ the passage between the connections is not unique---it depends on so-called Immirzi parameter \cite{immirzi}, which at the classical level labels a family of canonical transformations and thereby is physically irrelevant. It turns out, however, that there is no unitary implementation of these transformations at the quantum level. Thus one obtains a family of {\em inequivalent} models of LQG depending on the parameter. In particular, black hole entropy derived in LQG framework is known modulo the value of the parameter appearing as a factor in the formula describing the entropy \cite{bhe}; $(iii)$ the scalar constraint of GR expressed in terms of the $SU(2)$-connection is much more complicated than one written in terms of $SL(2,\C)$ one (see e.g. \cite{rev}). Consequently, the resulting scalar constraint operator \cite{qsd} is given by quite complicated and implicit formulae; there is a hope that applications of the $SL(2,\C)$-connection can simplify the form of the operator. In our opinion, this is sufficient motivation to look for solutions of both non-compactness and reality conditions problems.

In our previous paper \cite{oko} we discussed some attempts to construct a space of quantum states for a theory of  connections with a non-compact structure group i.e. attempts to solve the non-compactness problem. We concluded there that slight modifications of the construction known from theories of connections with a compact structure group may be insufficient to obtain a satisfactory result in the non-compact case. Thus to solve the problem one needs rather a radically different idea.
      
Such an idea was presented by Jerzy Kijowski in his review \cite{kpt-rev} on this author's Ph.D. thesis and was originally applied \cite{kpt} more than 25 years ago as an element of a procedure of quantization of a field theory---a procedure which does not depend either on the curvature of space-time or its global structure (in \cite{kpt} there was quantized a scalar field theory, for an application of the idea to electrodynamics see \cite{kpt-qed}, for an application to quantum scalar field theory on a lattice and quantum many-body systems see \cite{werp}). Before we will describe the idea in details let us say briefly that it consists in using {\em projective techniques} to built the space of quantum states instead of {\em inductive ones} which are used in the compact case.   

The goal of this paper is to apply the Kijowski's proposal to quantization of a diffeomorphism invariant theory of connections with a {\em non-compact} structure group. However, because of the lack of any experience with quantization of theories of this sort we do not dare to apply it to GR right now. Instead we will find a very simple 'toy example' of such a theory and will quantize it combining the Kijowski's idea with the standard methods of LQG.  

We emphasize that in this paper we {\em do not} present any ideas which could solve the problem of reality conditions.

The paper is organized as follows: Section 2 contains the definition and brief description of the (classical) 'toy theory', Section 3 is devoted for quantization of the theory, while in Section 4 we present a discussion of the results obtained in Section 3. Before we will turn to the definition of the 'toy theory' let us first describe reasons for which the present LQG methods fail in the non-compact case and present the Kijowski's proposal.

\subsection{Failure of inductive techniques in the non-compact case}

The canonical quantization procedure providing the LQG model can be easily generalized (as it was done in \cite{cq-diff}) to one applicable to any diffeomorphism invariant theory of connections with a {\em compact} structure group. However, this general procedure has not been extended yet to one applicable to theories with non-compact structure groups. The reason is that in order to build the space of quantum states the procedure employs some inductive techniques which do not work well when the structure group is non-compact. Now we are going to describe the problem in details\footnote{For an equivalent description in terms of measures on the configuration space of (generalized) connections see \cite{oko}.}. 

Let us consider a theory of connections  with a (compact or non-compact) structure group $G$ and with local degrees of freedom. Clearly, such a theory possesses an infinite number of degrees of freedom---the configuration space is the infinite dimensional space $\A$ of the connections defined on a spatial slice $\Sigma$ of the spacetime underlying the theory. Our task now is to find the space of kinematic\footnote{The term 'kinematic' refers here to the fact that in general the quantum states we are going to construct cannot be called physical since we expect the existence of some constraints---the physical quantum states are supposed to be extracted from kinematic ones by the Dirac procedure.} quantum states for the theory. Because of complexity of the configuration space we proceed in two steps: first $(i)$ we reduce the number of degrees of freedom to a finite one obtaining a {\em reduced configuration space} and construct the space of quantum states for this case, then $(ii)$ we combine the spaces obtained in the first step into the desired space of quantum states of the full theory.   

The reduction of the infinite number of the degrees of freedom  proceeds as follows (for more details see e.g. \cite{proj,cq-diff}). One fixes a graph\footnote{A graph  is here a collection of finite number of oriented edges. In particular one edge $e$ also form a graph $\{e\}$.} $\g$ embedded in $\Sigma$ and defines the following equivalence relation on $\A$: connections $A_1,A_2\in\A$ are said to be equivalent, $A_1\sim_\g A_2$, if for every edge $e$ of the graph $\g$    
\[
h_e(A_1)=h_e(A_2)
\] 
where $h_e(A)$ is the holonomy of (i.e. parallel transport defined by) the connection $A$ along $e$. Then
\begin{equation}
\A_\g:=\A/\sim_\g
\label{ag}
\end{equation}
is a finite dimensional space considered to be a reduced configuration space. An important fact is that  $\A_\g$ is a manifold isomorphic to $G^N$ \cite{al-hoop}, where $N$ is the number of the edges of $\g$, hence $\A_\g$ is {\em (non-)compact} if and only if $G$ is {\em (non-)compact}. Now, the space $\h_\g$ of kinematic states for the reduced case is the Hilbert space 
\begin{equation}
L^2(\A_\g,d\mu_\g),
\label{l2}
\end{equation}
where $d\mu_\g$ is a measure on $\A_\g$ induced by the Haar measure on the corresponding $G^N$. Thus the first step of the procedure is done.

Let us now consider the second step. An important fact here is that (under some technical assumption) graphs embedded in $\Sigma$ form a directed set $({\rm Gra},\geq)$. Thus the family $\{\h_\g\}$ of the Hilbert spaces is labeled by the directed set and it is very tempting to endow the family with the structure of an {\em inductive family} for the inductive limit 
\begin{equation}
\h:=\underrightarrow{\lim} \,\h_\g
\label{h}
\end{equation}
of such a family would also be a Hilbert space which could serve as the space of kinematic quantum states we are looking for. In order to transform the family $\{\h_\g\}$ into an inductive one we have to define, for every pair $\g'\geq\g$, a linear map $p_{\g'\g}:\h_\g\rightarrow \h_{\g'}$ preserving the scalar product. Moreover, for every triple of graphs such that $\g''\geq\g'\geq\g$ the corresponding maps have to satisfy the following consistency condition:
\[
p_{\g''\g}=p_{\g''\g'}\circ p_{\g'\g}.
\]   

So far we know how to construct the embeddings $\{p_{\g'\g}\}$ only in the case when the structure group $G$ is {\em compact}. In a particular case, when an edge $e$ does not intersect the edges of a graph $\g$ and $\g':=\g\cup \{e\}$ we have $\g'\geq\g$ and the embedding $p_{\g'\g}$ is of the following form:  
\begin{equation}
\h_\g\ni \Psi\mapsto p_{\g'\g}\Psi:=\Psi\ot I\in \h_{\g'},
\label{pgg}
\end{equation}
where $I$ is a constant function on $\A_{e}$ of value equal to $1$. The definition of $p_{\g'\g}$ is correct because $(i)$ $\A_{\g'}=\A_\g\times\A_{e}$ and $(ii)$ function $\Psi\ot I$ is square integrable on $\A_{\g'}$ for the space is compact. 

It is clear now that in the case of a non-compact structure group this construction has to break down since the function $I$ is not square integrable any longer. The question now is whether we can use any other function instead of $I$. 

Consider then a graph $\g$ and edges $\{e_i\}$ $(i=1,2,3)$ such that $e_3=e_1\circ e_2$ (where '$\circ$' means the composition of the edges) and each $e_i$ does not intersect the edges of $\g$. Denote
\[
\g_i:=\g\cup \{e_i\}, \ i=1,2,3 \ \ \ \text{and} \ \ \ \g_4:=\g\cup \{e_1\}\cup \{e_2\}.
\]
Note that the graph $\g_4$ can be obtained from $\g$ in two ways: $(i)$ by adding to $\g$ the edge $e_3$ in the first step (resulting in the graph $\g_3$) and by splitting the edge $e_3$ into $e_1$ and $e_2$ in the second step or $(ii)$  by adding the edge $e_1$ to $\g$ in the first step and by adding the edge $e_2$ to the just obtained graph $\g_1$ in the second one. Since the directing relation on the set ${\rm Gra}$ is defined in such a way (see e.g. \cite{cq-diff}) that $\g_4\geq\g_3\geq\g_i\geq\g$ the two ways of transforming $\g$ to $\g_4$ give rise to 
\begin{equation}
p_{\g_4\g_3}\circ p_{\g_3\g}=p_{\g_4\g_1}\circ p_{\g_1\g}.
\label{pp-pp}
\end{equation}

Let the map $p_{(\Gamma\cup\{e_i\})\Gamma}$ corresponding to adding the edge $e_i$ to a graph $\Gamma$ be of the following form
\[
\h_\Gamma\ni \Psi\mapsto p_{(\Gamma\cup \{e_i\})\Gamma}\Psi:=\Psi\ot \Psi_i\in \h_{\Gamma\cup \{e_i\}}
\]      
where $\Psi_i\neq 0$ is a square integrable function on ${\cal A}_{\{e_i\}}\cong G$ which is supposed to replace the function $I$ appearing in \eqref{pgg}. Some methods commonly used in LQG (see again \cite{cq-diff}) allow us to transform \eqref{pp-pp} to the following condition\footnote{The equation makes sense if we treat the functions $\{\Psi_i\}$ as ones on $G$.}
\begin{equation}
\Psi_3(g_1g_2)=\Psi_1(g_1)\Psi_2(g_2)
\label{P-PP}
\end{equation}
which has to be satisfied for every $g_1,g_2\in G$.

Now we restrict our considerations to edges $\{e_i\}$ which are pairwise diffeomorphic. Since our goal is the space of quantum states equipped with an action of diffeomorphisms of $\Sigma$ on it it is difficult to avoid an assumption that $\Psi_1=\Psi_2=\Psi_3$. This conclusion and Equation \eqref{P-PP} mean that $\Psi_3$ defines a homomorphism from $G$ into the {\em commutative} multiplicative group $\C\setminus\{0\}$ that is, a representation of $G$ on $\C$. Moreover, to define some operators (like e.g. area operator \cite{area}) on the space of quantum states we usually use derivations. Because of that it is reasonable to assume that the function $\Psi_3$, that is, the homomorphism is also differentiable. Now one can show that if $G$ is semi-simple (like e.g. $SL(2,\C)$ or $SU(1,1)$ appearing in $2+1$ gravity) then the homomorphism has to be trivial which means that $\Psi_3=I$.

It seems then that if we insist to construct the space of quantum states via the inductive techniques we have to use the function $I$, which means that the measure $d\mu_\g$ has to be (proportional to) a probability measure on $\A_\g$. The problem is that by now nobody was able to show that in the non-compact case $(i)$ there exist probability measures $\{d\mu_\g\}$ on $\{A_\g\}$ such that they would provide us with a space of quantum states equipped with a non-trivial unitary action of the Yang-Mills gauge group and the diffeomorphism group, which are symmetries of GR and that $(ii)$ there is a unique (or at least 'natural') choice of such a measure.       

Thus, in the (most interesting for us) case of semi-simple non-compact structure groups, we are not able to use the inductive limit to 'glue' the spaces $\{\h_\g\}$ into the desired Hilbert space\footnote{It turns out that in the non-compact case it is possible to use an orthogonal sum to 'glue' the spaces $\{\h_\g\}$ \cite{freidel} into a large Hilbert space, but then there exist obstacles \cite{oko} which do not allow us to define any acceptable representation of classical observables on the large Hilbert space.}.       

\subsection{Projective techniques}
 
The cornerstone of the Kijowski's proposal is the relation between graphs $\gamma'\geq\gamma$ seen as the relation {\em 'system-subsystem'}. Once $\gamma$ is recognized as a subsystem of $\gamma'$ the validity of application of the inductive techniques to the construction of the space of quantum states becomes not so obvious as before. Moreover, this new point of view suggests strongly that projective techniques should be used instead of the inductive ones. Kijowski explains his point of view as follows \cite{kpt-rev}:

\begin{quote}
 ``(...) to construct\footnote{The original text in Polish was translated into English by this author.} the Hilbert space by the inductive limit means to associate in a unique way every state of a {\em subsystem} (a small graph) with a state of a {\em system} (a large graph). This is not a natural procedure: in order to define such an embedding we have to choose arbitrarily a physical state corresponding to those degrees of freedom of the large system  which are neglected while defining its subsystem. If the configuration space encompassing all degrees of freedom is compact then the topology suggests a unique choice: the physical state should be described by a constant wave function on the neglected degrees of freedom. However, in a general case of a non-compact configuration space there is no natural choice of such a state\footnote{This statement is, in fact, very important since it gives the most general argument against applying inductive techniques in the non-compact case. Recall that in the previous subsection we showed that inductive techniques require the constant function on the structure group to be square integrable. This conclusion is not general, since it is based on the assumption that an edge can be expressed as a {\em composition} of two other edges. If one drops this assumption (e.g. by using so-called almost analytic loops introduced in Subsection \ref{s-s-ISQ}) then there is no way to get Equation \eqref{P-PP} which is indispensable for the conclusion.}. Thus, in my opinion, one should try to base the construction on projective limits. From the physical point of view it means that we associate in a unique way every state of the {\em system} with a (mixed, in general!) state of its {\em subsystem}---the latter state is obtained by 'forgetting' about the neglected degrees of freedom. Such a 'forgetting' operator is well defined and it was introduced for the first time in the well known analysis of the Einstein-Rosen-Podolsky 'experiment': a mixed state of one particle is obtained here from a two-particle state by 'forgetting' about the other particle. This procedure matches our basic physical intuition concerning the relation 'system-subsystem'.''   
\end{quote}

This suggests the following strategy of constructing the space of (kinematic) quantum states for a theory of connection with any (compact or non-compact) structure group: $(i)$ given graph $\g$, define the space ${\cal D}_\g$ of mixed states as the space of density matrices on the Hilbert space $\h_\g$, next  $(ii)$ by defining appropriate projections $\pi_{\g'\g}:D_{\g'}\rightarrow D_{\g}$ (i.e. 'forgetting' operators) endow the set $\{{\cal D}_\g\}$ with the structure of a projective family and, finally, $(iii)$ define the desired space $\D$ of kinematic quantum states as the {\em projective limit} of the family.      

It is important to note that the resulting space $\D$ of quantum states {\em is not} a Hilbert space---it is a {\em convex set} and can be interpreted as a space of {\em mixed} states of the quantized theory.

\section{Classical theory}

The present section is devoted for the definition and brief description of a classical theory of connections with a non-compact structure group---this theory will be quantized in the sequel of the paper. Since we do not have any experience in quantizing theories of this sort we would like to find a very simple example of such a theory. On the other hand, we would like to deal with an example related to general relativity (GR) as closely as possible. To satisfy the former wish  we will choose a theory whose structure group is the group $\R$ of real numbers  which seems to be the simplest non-compact Lie group. To fulfill the latter one we will define the theory by means of an action resembling the Pleba\'nski action for GR. 

We will begin this section by introducing the Lagrangian formulation of the theory. Then, since the quantization procedure we are going to apply is closely related to canonical quantization, we will describe the Hamiltonian framework of the theory including the algebra of constraints. Finally, we will briefly compare the theory to GR.
\subsection{Lagrangian formulation}

Let us consider a trivial principle bundle $P:={\cal M}\times \R$, where the set $\R$ of real numbers equipped with the addition plays the role of the (non-compact, commutative) structure group of the bundle, and the base manifold ${\cal M}$ is 4-dimensional and real-analytic. We assume also that there exists a 3-dimensional real analytic manifold $\Sigma$ such that ${\cal M}=\Sigma\times\R$.  

Let $A$ be a connection on the bundle $P$. Clearly, it can be represented by a real one-form on ${\cal M}$, which will be denoted by $A$ also. Consider now a theory, called in the sequel the {\em 'toy theory'}, given by the following action
\begin{equation}
S[A,\sigma,\Psi]:=\int_{\cal M} \sigma\wedge F -\frac{1}{2}\Psi\, \sigma\wedge\sigma,
\label{action}
\end{equation}      
where $F=dA$ is the curvature of the connection $A$, $\sigma$ is a two-form on ${\cal M}$ valued in the Lie algebra of $\R$ (which will be naturally identified\footnote{In this case the map $\exp$ from the Lie algebra into the group $\R$ is a bijection which provides us with the identification.} with $\R$) and $\Psi$ is a real valued function on the manifold.

\subsection{Hamiltonian formulation}

The action \eqref{action} can be expressed as
\begin{multline*}
S=\frac{1}{4}\int_{\cal M} (\sigma_{\alpha\beta} F_{\mu\nu} -\frac{1}{2}\Psi\sigma_{\alpha\beta}\sigma_{\mu\nu})\,dx^\alpha\wedge dx^\beta \wedge dx^\mu \wedge dx^\nu=\\=
\frac{1}{4}\int_{\cal M} (\sigma_{\alpha\beta} F_{\mu\nu} -\frac{1}{2}\Psi\sigma_{\alpha\beta}\sigma_{\mu\nu})\tilde{\epsilon}^{\alpha\beta\mu\nu}\,dx^4,
\end{multline*}
where $(x^\alpha)$, $(\alpha=0,1,2,3)$ are (local) coordinates on $\cal M$, and $\tilde{\epsilon}^{\alpha\beta\mu\nu}$ is the Levi-Civita density on $\cal M$. Taking advantage of the assumption ${\cal M}=\Sigma\times\R$ and treating $\Sigma$ as a 'space-like' slice of $\cal M$ and $\R$ as a 'time' we assume that the coordinates $(x^i)$ $(i=1,2,3)$ are (local) coordinates on $\Sigma$, while $x^0\equiv t$ is a coordinate on $\R$. Then 
\begin{equation}
S=\int_{\R}dt\int_{\Sigma} dx^3\, [ \tilde{E}^i\dot{A}_i -(-A_0\partial_i\tilde{E}^i-\frac{1}{2}\sigma_{0i}\tilde{\epsilon}^{ijk}F_{jk}+2\Psi\sigma_{0i}\tilde{E}^i)],
\label{action-H}
\end{equation}
where $\dot{A}_i:=\partial_t A_i$, $\tilde{\epsilon}^{ijk}:=\tilde{\epsilon}^{0ijk}$ is the Levi-Civita density on $\Sigma$ and 
\begin{equation}
\tilde{E}^i(t,x^l):=\frac{1}{2}\tilde{\epsilon}^{ijk}\sigma_{jk}(t,x^l)    
\label{E}
\end{equation}
is, for every fixed $t$, a vector density on $\Sigma$. 

Thus we obtain a Hamiltonian 
\begin{equation}
H[\tilde{E}^i, A_i, A_0, \sigma_{0i}, \Psi]=-\int_{\Sigma}dx^3\,(A_0\partial_i\tilde{E}^i+\frac{1}{2}\sigma_{0i}\tilde{\epsilon}^{ijk}F_{jk}-2\Psi\sigma_{0i}\tilde{E}^i). 
\label{ham-0}
\end{equation}
It follows from \eqref{action-H} that $\tilde{E}^i(t,x^j)$ is the momentum canonically conjugated to the configuration variable $A_i(t,x^j)$ as an $\R$-connection on $\Sigma$ (i.e. a connection on the trivial principle bundle $\Sigma\times \R$), while the variables $ A_0, \sigma_{0i}$ and $\Psi$ are just Lagrange multipliers. Thus the phase space of the theory is 
\[
{\cal P}:=\tilde{\cal E}\times\A,
\]
where $\tilde{\cal E}$ is the space of all vector densities on $\Sigma$ and $\A$ is the space of all $\R$-connections on $\Sigma$. The Poisson bracket between a pair $(\xi,\zeta)$ of functions on the phase space is of the standard form
\begin{equation}
\{\xi, \zeta \}\ =\ \int_\Sigma dx^3 \left(\frac{\delta \xi}{\delta A_i(x)}
\frac{\delta \zeta}{\delta \tilde{E}^i(x)}-\frac{\delta \xi}{\delta \tilde{E}^i(x)}\frac{\delta \zeta}{\delta A_i(x)}
\right).
\label{poisson}
\end{equation}

Evidently, the Hamiltonian \eqref{ham-0} is a sum of constraints. One of the constraints turns out to be easily solvable. Indeed, the variation of the Hamiltonian with respect to $\Psi$ gives us
\[
\sigma_{0i}\tilde{E}^i=0.
\]  
The general solution of the equation is
\[
\sigma_{0i}=\,^{\text{\tiny{--1}}}\!\tilde{\epsilon}_{ijk}\,N^j\tilde{E}^k,
\]
where $\,^{\text{\tiny{--1}}}\!\tilde{\epsilon}_{ijk}$ is the Levi-Civita density on $\Sigma$ of weight $-1$, and  $N^i(t,x^j)$ is, for every fixed $t$, a vector field  on $\Sigma$. Setting the solution to \eqref{ham-0} we obtain
\begin{equation}
H[\tilde{E}^i, A_i, A_0, N^i]=-\int_{\Sigma}dx^3\,(A_0\partial_i\tilde{E}^i+N^i\tilde{E}^jF_{ij}),
\label{ham}
\end{equation}
where $N^i$ is a Lagrange multiplier, $F_{ij}$ is the curvature two-form of $A_i$ while the role of the other variables remains unchanged. 

The quantization of the 'toy theory' will be based on the Hamiltonian \eqref{ham}.

\subsection{Constraints and gauge transformations}

The Hamiltonian \eqref{ham} is a sum of the following constraints:
\begin{equation}
\begin{aligned}
&{\cal C}_{\rm G}(\tilde{E}^i,A_j):=\partial_i\tilde{E}^i,\\
&{\cal C}_k(\tilde{E}^i,A_j):=\tilde{E}^jF_{kj}.
\end{aligned}
\label{constr}
\end{equation}
In canonical theory constraints play a double role \cite{teitelboim}: $(i)$ they define a physical subspace of the phase space of the theory and $(ii)$ generate gauge transformations on the phase space. Here the physical subspace is given by the equations  
\begin{equation}
{\cal C}_{\rm G}(\tilde{E}^i,A_j)=0, \ \ \ \text{and} \ \ \ {\cal C}_k(\tilde{E}^i,A_j)=0
\label{ph-phase},
\end{equation}
which possess the following large class of solutions 
\begin{equation}
\{(\tl{E}^i(A_j,g),A_j)\}
\label{con-sol}
\end{equation}
where
\begin{equation}
\tl{E}^i(A_j,g)=g \,\tl{\epsilon}^{ijk}F_{jk},
\label{sol-E}
\end{equation}
and $g$ is a smooth function on $\Sigma$ such that
\begin{equation}
\tl{\epsilon}^{ijk}F_{jk}\partial_i g=0
\label{sol-g}.
\end{equation}
 
It is easy to give a natural interpretation for the gauge transformations resulting from the constraints \eqref{constr} if they are combined into functions
\begin{align}
C_{\rm G}(\Omega)&:=\int_{\Sigma}dx^3\,\Omega\,{\cal C}_{\rm G}(\tilde{E}^i,A_j),\ \ \ \Omega:\Sigma\rightarrow \R \label{gauss},\\
C_{\rm Diff}(\vec{N})&:=\int_{\Sigma}dx^3\,[N^k\,{\cal C}_k(\tilde{E}^i,A_j)-(N^iA_i)\,{\cal C}_{\rm G}(\tilde{E}^i,A_j)] \label{diff}.
\end{align}
Then the gauge transformation  are described by the following differential equations, respectively,
\begin{align*}
\frac{d}{d\tau} A_i&=\{A_i,C_{\rm G}(\Omega)\}=-\partial_i \Omega,\\
\frac{d}{d\tau} \tilde{E}^i&=\{\tilde{E}^i,C_{\rm G}(\Omega)\}=0.
\end{align*}
and
\begin{align*}
\frac{d}{d\tau} A_i&=\{A_i,C_{\rm Diff}(\vec{N})\}=({\cal L}_{\vec{N}}A)_i,\\
\frac{d}{d\tau} \tilde{E}^i&=\{\tilde{E}^i,C_{\rm Diff}(\vec{N})\}=({\cal L}_{\vec{N}}\tilde{E})^i,
\end{align*}
where ${\cal L}_{\vec{N}}$ denotes the Lie derivative along the vector field $\vec{N}$. Integrating the above formulas we obtain, respectively,
\begin{equation}
A^{(\tau)}=A-\tau d\Omega, \ \ \ \ \tilde{E}^{(\tau)}=\tilde{E},
\label{y-m}
\end{equation}
and
\begin{equation}
A^{(\tau)}=\chi_{-\tau}^* A, \ \ \ \ \tl{E}^{(\tau)}=\chi_{\tau*}\tl{E},
\label{diff-ac}
\end{equation}
where $\chi_\tau$ is an element of the one-parameter family of diffeomorphisms on $\Sigma$ generated by  the vector field $\vec{N}$. We easily recognize the former transformations as a Yang-Mills gauge transformation generated by the function $\Omega$ , and the latter ones---as an action  of the diffeomorphism. Thus it is justified to call $C_{\rm G}(\Omega)$ the Gauss constraint, and $C_{\rm Diff}(\vec{N})$---the diffeomorphism one.    

The Poisson bracket \eqref{poisson} between functions \eqref{gauss} and \eqref{diff} are of the following form
\begin{equation}
\begin{aligned}
\{C_{\rm G}(\Omega), C_{\rm G}(\Omega')\}&=0\\
\{C_{\rm G}(\Omega), C_{\rm Diff}(\vec{N})\}&=-C_{\rm G}({\cal L}_{\vec{N}}\Omega),\\
\{C_{\rm Diff}(\vec{N}), C_{\rm Diff}(\vec{M})\}&=C_{\rm Diff}([\vec{N},\vec{M}]),
\end{aligned}
\label{const-alg}
\end{equation}
This means that the constraints are of the first class and, for the Hamiltonian is the sum of the constraints, each of them is preserved by the time evolution.

Let us finally comment on the physical degrees of freedom of the 'toy theory'. Before the constraints \eqref{ph-phase} are solved there are $6$ degrees of freedom per point in $\Sigma$. Since there are $4$ constraints then taking into account the gauge transformations the number of true physical degrees of freedom seems to be reduced to $6-2\cdot4=-2$ per point which sounds worrisome\footnote{This issue, overlooked originally by this author, was pointed out to him by Prof. Abhay Ashtekar.}. A more detailed analysis (see Appendix \ref{deg-free}) shows however, that the 'toy theory' possesses an uncountable number of physical degrees of freedom (the conclusion is true at least in the case when $\Sigma=\R^3$).

\subsection{The 'toy theory' versus general relativity \label{versus}}

The form of the action \eqref{action} is fully analogous to the form of the well known Pleba{\'n}ski self-dual action \cite{pleb}:
\begin{equation}
S[A^{A}\!_{B},\Sigma^{A}\!_{B},\Psi_{ABCD}]=\int_{\cal M} \Sigma^{AB} \wedge F_{AB} -
\frac{1}{2}\Psi_{ABCD}\,\Sigma^{AB} \wedge\Sigma^{CD},
\label{pl}
\end{equation}
where $F^{A}\!_{B}$ (the index\footnote{To raise and lower the indices one uses the antisymmetric bilinear form $\epsilon_{AB}$.} $A=0,1$) is the curvature form of an $SL(2,\C)$-connection $A^{A}\!_{B}$ on ${\cal M}$, $\Sigma^{A}\!_{B}$ is a two-form valued in the Lie algebra of $SL(2,\C)$, and $\Psi_{ABCD}=\Psi_{(ABCD)}$ is a symmetric spinor field on the manifold. Performing the Legendre transformation and solving the constraint 
\[
\Sigma^{(AB}_{0i}\tilde{E}^{iCD)}=0,\ \ \ \tilde{E}^{iA}\!_{B}:=\tilde{\epsilon}^{ijk}\Sigma_{jk}\!^{A}\!_{B}
\]
one obtains a Hamiltonian \cite{ashtekar,capov}:
\begin{multline}
H[\tilde{E}^{iA}\!_B,A_i\!^A\!_B,A_{0}\!^{A}\!_{B},N^{i},\,^{\text{\tiny{--1}}}\!\tilde{N} ]=\\=-\int_{\Sigma}dx^3\,  [\, A_{0}^{A}\!_{B}(D_{i}\tilde{E}^{iB}\!_{A})+ N^{i}\tilde{E}^{jA}\!_{B}F_{ij}^{B}\!_{A}+\,^{\text{\tiny{--1}}}\!\tilde{N}\tilde{E}^{iA}\!_{B}\tilde{E}^{jB}\!_{C}F_{ij}^{C}\!_{A}\, ],
\label{ham-grav}
\end{multline}
where
\begin{equation}
D_{i}\tilde{E}^{iA}\!_{B}:=\partial_i\tilde{E}^{iA}\!_B+A_i\!^A\!_C\tilde{E}^{iC}\!_B-A_i\!^C\!_B\tilde{E}^{iA}\!_C.
\label{cov-der}
\end{equation}
The canonical variables called {\em complex Ashtekar variables} are the $SL(2,\C)$-connection $A_i\!^A\!_B$ on $\Sigma$ as the configuration variable and  the vector density $\tilde{E}^{iA}\!_B$ valued in the Lie algebra of $SL(2,\C)$ as the momentum, while the time component $A_{0}\!^{A}\!_{B}$ of the connection form $A^A\!_B$, the vector field $N^{i}$ on $\Sigma$ and the density $ \,^{\text{\tiny{--1}}}\!\tilde{N}$ on $\Sigma$ of weight $-1$ are Lagrange multipliers.

The differences between the forms of the Hamiltonians \eqref{ham} and \eqref{ham-grav} originate merely in the differences between  the structure groups $\R$ and $SL(2,\C)$. More precisely, let $\tl{\R}$ be a subgroup of $SL(2,\C)$ isomorphic to $\R$. We restrict
\begin{itemize}
\item the phase space of GR to fields $(\tl{E}^{iA}\!_B,A_j\!^C\!_D)$ valued in the Lie algebra of $\tl{\R}$; 
\item the gauge group $SL(2,\C)$ to $\tl{\R}$.
\end{itemize}
In this way we obtain a theory describing $1+1$ {\em degenerate} sector of GR \cite{jac}. 

Note that after the restriction the Ashtekar variables can be expressed as 
\[
\tl{E}^{iA}\!_B=\tl{E}^{i}M^{A}\!_B \ \ \ \text{and} \ \ \ A_j\!^C\!_D=A_jM^C\!_D,
\]
where $M^{A}\!_B$ is a non-zero vector in the Lie algebra of $\tl{\R}$. It is clear that the two last terms in \eqref{cov-der} reduce now to zero and the covariant derivative becomes equal to a usual one. Moreover, the scalar constraint vanishes:
\[
\tilde{E}^{iA}\!_{B}\tilde{E}^{jB}\!_{C}F_{ij}^{C}\!_{A}=\tilde{E}^i\tilde{E}^{j}F_{ij}\,M^{A}\!_{B}M^{B}\!_{C}M^{C}\!_{A}=0.
\]
Assuming that $M^{A}\!_{B}M^{B}\!_{A}=1$ we see that the Hamiltonian \eqref{ham-grav} reduces to \eqref{ham}.

We conclude that the 'toy theory' describes $1+1$ degenerate sector of GR.

\section{Quantization of the 'toy theory' \label{quant-sc}}

Combining ideas of \cite{kpt} with the methods of LQG we propose the following sequence of steps as a strategy of quantization of the 'toy theory':
\begin{enumerate}
\item first, we will introduce a method aimed at reducing the degrees of freedom of the theory. 
\begin{enumerate}
\item we will begin by constructing a Lie algebra of elementary classical variables obtaining as a result the so-called Ashtekar-Corichi-Zapata (ACZ) algebra $\sca$ \cite{acz} consisting of some (Yang-Mills gauge invariant) functions on $\A$ and 'momentum' (flux) operators. This algebra will correspond to Yang-Mills gauge invariant functions on the phase space $\cal P=\tilde{\cal E}\times\A$ of the theory and will encompass all relevant degrees of freedom.   
\item the reduction will consist in distinguishing a subalgebra $\sca_{\la}\subset\sca$ generated by some smooth functions on a reduced configuration space $\A_{\tL}$ and 'momentum' operators ($\A_{\tL}$ will be defined analogously to the reduced configuration space $A_\g$ given by \eqref{ag}). We will call the algebra $\sca_\la$ a reduced classical system.  
\end{enumerate}
The result of this step will be a family $\{{\sca}_\la\}_{\la\in\Lambda}$ of reduced classical systems labeled by a directed set $(\Lambda,\geq)$.
\item in this step we will define the relation 'system-subsystem' among members of the family $\{{\sca}_\la\}_{\la\in\Lambda}$: we will observe that if $\la'\geq\la$ then ${\sca}_\la\subset{\sca}_{\la'}$ and consequently we will call $\sca_\la$ a subsystem of $\sca_{\la'}$. This step will also contain a detailed analysis of the relation whose conclusions will be necessary for the further steps of the procedure.
\item then we will quantize canonically every system ${\sca}_\la$ obtaining a quantum system $\S_\la$ which should be considered as a 'reduced'  quantum 'toy theory'. This step will be done as follows:  we will extend the Lie algebra $\sca_\la$ to an operator $*$-algebra $\hat{\sca}_\la$, then we  will define a representation of $\hat{\sca}_\la$ on a Hilbert space ${\h_\la}$, whose vectors will represent pure states of ${\cal S}_\la$. This will provide us with a space ${\cal D}_\la$ of mixed states of the quantum system represented by density matrices on $\h_\la$. By the very construction, the states in $\D_\la$ will be invariant under the Yang-Mills gauge transformations; 
\item next, we will organize the systems $\{\S_\la\}$ into a large quantum system $\S$ corresponding to the infinite dimensional phase space $\cal P$ quotiented by the Yang-Mills gauge transformations \eqref{y-m}:
\begin{enumerate}
\item  for every pair $\la'\geq\la$ we will construct a projection $\pi_{\la\la'}:{\cal D}_{\la'}\rightarrow\D_{\la}$ (that is, the 'forgetting' operator) promoting the family $\{\D_\la\}$ into a projective one $\{\D_\la,\pi_{\la\la'}\}$. The projective limit $\D$ of the family will give us the desired space of (Yang-Mills gauge invariant) quantum states for the large system $\S$; 
\item given $\la$, elements of $\D_\la$ define positive linear functionals on the $C^*$-algebra ${\cal B}_\la$ of bounded operators on $\h_\la$ \footnote{$\D_\la$ does not coincide with the space of {\em all} states on ${\cal B}_\la$ (i.e. $\D_\la$ form a set of {\em normal} states), however the states in $\D_\la$ separate points in ${\cal B}_\la$.}. This implies the existence of dual maps $\pi^*_{\la\la'}:{\cal B}_\la\rightarrow{\cal B}_{\la'}$ such that $\{{\cal B}_\la,\pi^*_{\la\la'}\}$ is an inductive family. Its inductive limit $\cal B$ will play the role of the algebra of quantum observables, which can be evaluated on the states in $\D$.        
\end{enumerate}
Thus the quantum system $\S$ will consist of the space $\D$ and the $C^*$-algebra $\cal B$.     
\item finally, we will show that there exist an action of the diffeomorphisms of $\Sigma$ on the space $\D$ which naturally corresponds to the action \eqref{diff-ac} of the diffeomorphism constraint on the phase space. The space $\D_{\rm ph}$ of physical states of the 'toy theory' will be defined as the set of diffeomorphism invariant elements of $\D$. The result of this step will be a quantum system $\S_{\rm ph}:=(\D_{\rm ph},{\cal B})$ considered as the quantum  'toy theory'. 
\end{enumerate}

It turns out that in the particular case of the 'toy theory' the algebras $\{\sca_\la\}$ i.e. the reduced classical systems can be defined in two different ways: they can  be obtained by reduction of either $(i)$ only configuration degrees of freedom or $(ii)$ both momentum and configuration ones. In the first case the algebra $\sca_\la$ will describe {\em infinite} number of degrees of freedom, while in the second one---{\em finite} one. Quantization based on the first method as a simpler one will be presented in this paper, while quantization based on the second one will be described in the accompanied paper \cite{diff-II}. 
   
Let us emphasize that the quantum 'toy theory' will consist merely of the $C^*$-algebra $\cal B$ of quantum observables and the space $\D_{\rm ph}$ of diffeomorphism invariant states on it without any Hilbert space. In principle, one can obtain a Hilbert space for this theory once a state in $\D_{\rm ph}$ is distinguished---it is enough then to use the GNS construction to get a representation of $\cal B$ on a Hilbert space as it was done in \cite{kpt}. However, in the case of the 'toy theory' it seems to be no natural way to single out any state\footnote{The theory analyzed in \cite{kpt} possesses a non-vanishing Hamiltonian, which can be used to single out a state of minimal energy; in the 'toy theory' Hamiltonian is a sum of constraints and every state in ${\cal D}_{\rm ph}$ is annihilated by it.} in ${\cal D}_{\rm ph}$.

\subsection{Step 1: Reduction of degrees of freedom}

To obtain the full information about a field configuration we need to measure an infinite number of quantities corresponding to the degrees of freedom of a theory, hence we need an infinite number of measuring instruments. Thus, in general, the reduction of degrees of freedom is performed by restricting oneself to some (usually finite) number of the instruments. Let us then begin Step 1 of the quantization procedure by defining the instruments.

\subsubsection{Measuring instruments}

Let $\varphi$ and $h$ be instruments which can be used to measure values of the fields, respectively, $\tilde{E}$ and $A$ on $\Sigma$. We assume also that the instruments are imperfect in the sense that their sensitivity varies from point to point---the numerical values of corresponding measurements are given by \cite{kpt}
\[
\varphi(\tilde{E})=\int_\Sigma \tilde{E}^i\varphi_{i}\ \ \ \text{and}\ \ \ h(A)=\int_\Sigma A_i\,\tl{h}^i,
\]
where $\varphi=\varphi_{i}\,dx^i$ is a differential one-form on $\Sigma$, and $\tl{h}^i$ is a vector density (in general, both $\varphi$ and $\tl{h}$ can be of distributional character).     

The class of instruments, as defined above, is too broad for quantization of our theory. Following LQG methods \cite{ai,area,acz} we restrict ourselves to instruments described by the following formulas
\begin{equation}
\begin{aligned}
\varphi_{S,f}(\tilde{E})&=\int_{S} *\tl{E}f,\\
h_{l}(A)&=\int_{l} A,
\end{aligned}
\label{instr}
\end{equation}
where $S$ is a bounded {\em oriented} surface (two-dimensional submanifold) in $\Sigma$, $*\tl{E}$ is a two-form on $\Sigma$ defined by the Levi-Civita form $^{\text{\tiny{--1}}}\tl{\epsilon}_{ijk}$ of weight $-1$,
\[
*\tl{E}:=\tl{E}^i\ ^{\text{\tiny{--1}}}\tl{\epsilon}_{ijk}\,dx^j\wedge dx^k,
\]
$f$ is a real (smooth) function on $S$, and $l$ is an oriented loop embedded in $\Sigma$. The quantities $\varphi_{S,f}(\tl{E})$ and $h_l(A)$ are nothing else but, respectively, a flux of the field $\tl{E}$ across the surface $S$ \cite{area} and the holonomy of the connection $A$ along the loop $l$ \cite{oko}. The latter one obviously satisfies 
\begin{equation}
h_{l\circ l'}=h_l+h_{l'}
\label{h-lin}
\end{equation} 
where $l\circ l'$ is the composition of the loops $l$ and $l'$.      

It turns out that, in order to force the quantization machinery to work, we have to impose some rather technical assumptions  on the manifold $\Sigma$, the surface and the loop. We assume then that
\begin{enumerate}
\item $\Sigma$ is a real analytic manifold; 
\item the surface $S$ is an analytic submanifold of $\Sigma$; 
\item the loop $l$ is piecewise analytic and bases at an arbitrary but fixed point\footnote{As it will be shown later, no particular choice of the point $y$ effects the quantization procedure (see the discussion at the beginning of Subsection \ref{sol-diff}).} $y\in\Sigma$.
\end{enumerate} 
These assumptions are common in the LQG literature (see e.g. \cite{al-hoop, area, acz})---they will allow us to define the ACZ algebra $\sca$. However, the quantization method we are going to apply requires stronger assumptions\footnote{The assumptions just presented are sufficient for the other quantization method which will be described in \cite{diff-II}.} which will be imposed on the measuring instruments later on. 

One can easily realize that two different loops can define the same instrument $h$. To obtain an unambiguous labeling of this sort of instruments one introduces the notion of a {\em hoop} \cite{al-hoop}. Denote by $\mathbb{L}_y$ the set of all piecewise analytic loops based at $y$ and consider the set of the instruments $\{h_{l}\}$ given by all the elements of $\mathbb{L}_y$. Now, two loops $l,l'\in \mathbb{L}_y$ are said to be {\em holonomically equivalent}, $l\sim l'$, if and only if for every connection $A\in\A$:
\[
h_l(A)=h_{l'}(A).
\]  
Denote by $\tll$ the equivalence class of loop $l$ and call it a {\em hoop}. The set of all hoops
\[
{\cal HG}:=\mathbb{L}_y/\sim
\] 
is an Abelian group called the {\em hoop group} with the group action given by
\[
\tll\circ\tll'=\widetilde{l\circ l'},
\]  
where $l\circ l'$ is a composition of the loops $l$ and $l'$. Clearly, there is one-to-one correspondence between elements of $\h\G$ and the instruments $\{h_{l}\}$. Since now we will then use the symbol $h_{\tll}\equiv h_{l}$ also. 

Note finally that all the measuring instruments under consideration produce {\em Yang-Mills gauge invariant} outcomes which easily follows from \eqref{y-m}---this is a reason why we have used loops to define the instrument $\{h_l\}$ instead of edges (of a graph).

\subsubsection{ACZ algebra}

In the present subsection we are going to define the algebra of elementary variables encompassing all the (Yang-Mills gauge invariant) degrees of freedom. From the technical point of view our task consists merely in adapting the original construction of ACZ-algebra \cite{acz} to the 'toy theory'.

Obviously, what we call measuring instruments are nothing else but functions on the phase space of the 'toy theory'. Therefore we can calculate a Poisson bracket between the instruments \cite{acz,area}:
\begin{align}
&\{h_{l_I},h_{l_J}\}=0=\{\varphi_{S_I,f_I},\varphi_{S_J,f_J}\},\label{hhff-bracket}\\
&\{\varphi_{S,f},h_l\}=-\frac{\check{f}}{2}, \label{f-h-bracket}
\end{align}
where $\check{f}$ is a real number (a method of finding it will be presented below). Let us also emphasize that the vanishing of the Poisson bracket between $\varphi_{S_I,f_I}$ and $\varphi_{S_J,f_J}$ (Equation \eqref{hhff-bracket}) is not a na\"{\i}ve conclusion drawn from the fact that both functions depend only on the momentum variable $\tl{E}$---the vanishing of the bra\-cket is implied in fact by the commutativity of the structure group $\R$ as it can be shown by means of a rigorous method developed in \cite{acz} (see also the remark below Definition \ref{acz-df}). 

To obtain the number $\check{f}$ we subdivide the loop $l$  on a finite number of connected and oriented segments (the orientation of the segment is inherited from the orientation of the loop) such that each segment is either $(i)$ contained in $S$ (modulo its endpoints) or $(ii)$ the intersection of the segment with $S$ coincides with precisely one endpoint of the segment or $(iii)$ the segment does not intersect\footnote{The subdivision is possible thanks to analyticity of $l$ and $S$---if the loops and the surface were only smooth they could have an infinite number of intersection points which would make the bracket \eqref{f-h-bracket} ill defined.} $S$. Let
\[
f^+:=\sum_k f(y_k),
\]
where  $\{y_k\}$ are all the intersection points between $S$ and those segments of the kind $(ii)$ which either are 'outgoing' from $S$ and placed 'above'\footnote{The term {'}'above' ('below') the surface' refers to the orientation of $S$.} the surface or are 'incoming' to $S$ and are placed 'below' the surface. Similarly, 
\[
f^-:=\sum_n f(y_n),
\]
where $\{y_n\}$ are all the intersection points between $S$ and those segments of the kind $(ii)$ which either are 'outgoing' from $S$ and are placed 'below' the surface or are 'incoming' to $S$ and are placed 'above' the surface. Then
\[
\check{f}:=f^+-f^-.
\]   

Now we are going to define a kind of cylindrical functions, which will be used to define the ACZ algebra $\sca$. Let us begin by recalling the definition of a {\em tame subgroup} of $\cal HG$ \cite{al-hoop,proj}.

\begin{df}
A finite subset ${\cal L}=\{l_1,\ldots,l_M\}$ of $\mathbb{L}_y$ is called a set of independent loops if and only if  {\em (i)} each loop $l_I$ contains an open segment which is traversed only once and which is shared by any other loop at most at a finite number of points and {\em (ii)} it does not contain any path of the form $e\circ e^{-1}$ where $e$ is a piecewise analytic path in $\Sigma$.
\label{ind-loops}
\end{df}

Let ${\cal L}=\{ l_1,\ldots, l_M\}$ be a set of independent loops. A subgroup  $\tl{\cal L}$ of the hoop group $\cal HG$ generated by hoops $\{\tll_1,\ldots,\tll_M\}$ is said to be a {\em tame subgroup} of $\cal HG$. In the sequel we will also say that the group $\tl{\cal L}$ is generated by the loops ${\cal L}$  which is not too precise but convenient. Given tame group $\tL$, there exist in general many holonomically inequivalent sets of independent loops such that each of them generates $\tL$.   
 
Given tame subgroup $\tL$, we define the following equivalence relation on the space $\A$ of the $\R$-connections on $\Sigma$: we say that connections  ${A}_1,A_2\in\A$ are equivalent, ${A}_1\sim_{\tL} {A}_2$, if and only if
\[
h_{\tll}(A_1)=h_{\tll}(A_2)
\]
for every hoop $\tll\in\tl{\cal L}$ \cite{al-hoop}. Then
\[
\A_{\tL}:=\A/\sim_{\tL}
\]
is a finite dimensional space. We will denote by $\pr_{\tL}$ the natural projection from $\A$ onto $\A_{\tL}$,
\begin{equation}
A\mapsto\pr_{\tL}(A):=[A],
\label{proj-l}
\end{equation}
where $[A]$ is the equivalence class of the connection $A$.  Evidently, the space $\A_{\tL}$ is an analogue of the reduced configuration space $\A_\g$ \eqref{ag} considered earlier. The following lemma describes properties of $\A_{\tL}$. 
\begin{lm}
Suppose that $\tL$ is generated by independent loops ${\cal L}=\{l_1,\ldots,$ $l_M\}$. Then the map
\begin{equation}
\A_{\tL}\ni [A]\mapsto {\cal I}_{\cal L}([A]):=(\,h_{l_1}(A),\ldots, h_{l_M}(A)\,)\in\R^M
\label{bij}
\end{equation}
is a bijection \cite{al-hoop} which equips $\A_{\tL}$ with a structure of a linear space. A linear structure on $\A_{\tL}$ given by a map ${\cal I}_{\cal L'}$, where $\cal L'$ is another set of   independent loops generating $\tL$, coincides with that defined by $\cal L$ i.e. the map ${\cal I}_{\cal L}\circ{\cal I}^{-1}_{\cal L'}:\R^M\rightarrow\R^M$ is linear and invertible \cite{oko}. 
\label{A=R}  
\end{lm}

Now we are going to recall the definition of Schwarz functions which will be used to define a kind of cylindrical functions. Let $\bld{\a}$ denote a multi-label $(\a_1,\ldots,\a_n)$ such that every $\alpha_i$ belongs to $\N=\{0,1,2,\ldots\}$. Given a smooth function $\psi:\R^N\rightarrow\C$, we denote by $D^{\bld{\a}}$ a partial derivative
\[
D^{\bld{\a}}\psi:=\frac{\partial^{\sum_i \a_i}}{\partial x^{\a_1}_1\ldots \partial x^{\a_n}_n} \psi,
\] 
where $(x_i)$ are the Cartesian coordinates on $\R^N$. A smooth function $\psi:\R^N\rightarrow\C$ is said to be  a Schwarz function  if and only if for every $m\in \N$ and for every derivative $D^{\bld{\a}}\psi$ 
\[
\lim_{r\rightarrow\infty} {(D^{\bld{\a}}\psi)}\,{r^{m}}=0,
\] 
where $r=\sqrt{\sum_i x^2_i}$. It is clear that Schwarz functions on $\R^N$ form a $*$-algebra which will be denoted by $\mathbb{S}_N$. 

We will also need another kind of functions defined on $\R^N$:
\begin{df}
We say that a smooth function $\psi:\R^N\mapsto\C$ is a multiplier of the $*$-algebra $\mathbb{S}_N$ if and only if for every derivative $D^{\bld{\a}}\psi$ there exist finite set $\{P_{1,\bld{\a}},\ldots,P_{n_{\bld{\a}},\bld{\a}}\}$ of polynomials on $\R^N$ such that
\[
|D^{\bld{\a}}\psi|\leq \sum_{i=1}^{n_{\bld{\a}}} |P_{i,\bld{\a}}|.
\]     
\label{mult-df}
\end{df}
We will denote the set of all multipliers of $\mathbb{S}_N$ by $\mathbb{M}_N$. It is a simple exercise to show that $(i)$ for every Schwarz function $\psi$ on $\R^N$ and for every multipliers $\psi'$ of $\mathbb{S}_N$ the product $\psi\psi'$ is again Schwarz function  and $(ii)$ $\mathbb{M}_N$ is a unital $*$-algebra.     

\begin{df}
The set $\cyl_{\tL}$ $(\cyl^S_{\tL})$ of cylindrical functions (Schwarz cyli\-ndrical functions) compatible with the tame group ${\tL}$ is the set of all complex functions on $\A$ of the form
\[
\Psi=\pr_{\tL}^*\,{\cal I}_{\cal L}^*\,\psi,
\]    
where $\psi$ is any element of $\mathbb{M}_N$ $(\mathbb{S}_N)$, $\pr_{\tL}$ is the projection \eqref{proj-l} and ${\cal I}_{\cal L}$ is the map \eqref{bij}.   
\label{cyl-df}
\end{df}
By virtue of Lemma \ref{A=R} the spaces $\cyl_{\tL}$ and $\cyl^S_{\tL}$ do not depend on the choice of the set $\cal L$ of independent loops generating the tame group $\tL $. Both spaces are $*$-algebras, and  $\cyl_{\tL}$ possesses a unit given by the function on $\R^N$ of constant value equal to $1$. 

Consider now a (complex) vector space $\Cyl$ spanned by all the cylindrical functions,
\begin{equation}
\Cyl:={\rm span}\{\ \cyl_{\tL}\ | \ \tL\subset{\cal HG}\ \}
\label{cyl-alg}
\end{equation}
and a set $\Phi$ of all operators on $\Cyl$ given as finite linear combinations of operators of the form
\begin{equation}
\phi_{S,f}\Psi:=\{\varphi_{S,f},\Psi\},
\label{hat-phi}
\end{equation}
where $S$ and $f$ run through all admissible surfaces and functions. To convince ourselves that the definition of $\Phi$ is correct we have to check whether every $\{\varphi_{S,f},\Psi\}$ is an element of $\cyl$. In fact, every operator \eqref{hat-phi} preserves each space $\cyl_{\tL}$. Indeed, let $\tL$ be generated by the set ${\cal L}=\{l_1,\ldots,l_N\}$ of independent loops. Then by virtue of \eqref{f-h-bracket} we get
\begin{equation}
\{\varphi_{S,f},\Psi\}=-\frac{1}{2}\,\pr_{\tL}^*\,{\cal I}_{\cal L}^*\, [\, \sum_{J=1}^N \check{f}_{J}\frac{\partial \psi}{\partial x^J}\,],
\label{e-psi-braket}
\end{equation}
where  
\[
\check{f}_J:=-2\{\varphi_{S,f},h_{l_J}\}
\]
and $(x^J)$ are the canonical coordinates on $\R^N$ (see \eqref{bij}). In fact, the r.h.s. of the above equation belongs to $\cyl_{\tL}$ for the algebra $\mathbb{M}_N$ is preserved by any derivative.

\begin{df}
The Ashtekar-Corichi-Zapata algebra \cite{acz} is a complex vector space 
\[
\sca:=\Cyl\times\Phi
\]
equipped with the Lie bracket 
\[
[(\Psi,\phi),(\Psi',\phi')]:=(\phi\Psi'-\phi'\Psi,[\phi,\phi']).
\]  
\label{acz-df}
\end{df}

The commutativity of $\R$ implies
\[
[\phi,\phi']=0
\] 
for every $\phi,\phi'\in\Phi$. This is, in fact, why the second equation of \eqref{hhff-bracket} holds for
\[
\{\varphi_{S_I,f_I},\varphi_{S_J,f_J}\}:=[\phi_{S_I,f_I},\phi_{S_J,f_J}].
\] 

\subsubsection{Reduced classical systems \label{cs}}

In the previous subsection we have defined the ACZ algebra $\sca$ encompassing all Yang-Mills gauge invariant degrees of freedom of the classical theory. Now we are ready to perform the reduction of the degrees of freedom.

As mentioned earlier we will distinguish a reduced classical system by limiting our attention to outcomes provided by a restricted set of measuring instruments. This can be done by distinguishing a subalgebra of $\sca$ generated by
\begin{enumerate}
\item an algebra of cylindrical function $\cyl_{\tL}$, 
\item a linear subspace $F$ of $\Phi$.
\end{enumerate}
In this paper we choose trivially $F=\Phi$ \footnote{Reduced systems obtained by choosing $F$ as a proper subspace of $\Phi$ will be described in \cite{diff-II}.}:  
\begin{df}
A subalgebra
\[
\sca_{\tL}=\Phi\times\cyl_{\tL}
\]
of $\sca$ is called a reduced classical system. 
\label{ics}
\end{df}

\subsubsection{The Kijowski's definition of reduced classical systems versus the pre\-sent ones \label{versus-sec}}

While quantizing a scalar field theory Kijowski \cite{kpt} reduces the infinite dimensional phase space of the theory to a finite dimensional one by restricting his attentions to outcomes provided by a finite number of measuring instruments which (from mathematical point of view) are defined as some integrals over three-dimensional regions. Following the LQG methods we decided to use instruments which are distributional in that sense that they are defined as integrals along curves and surfaces. This choice (as it is shown in \cite{acz}) leads naturally to the ACZ algebra $\sca$ which replaces the phase space $\cal P$ with its Poisson structure. This is the main reason why we defined reduced classical systems as subalgebras of the ACZ algebra $\sca$ instead of reduced phase spaces, though each of the subalgebras can be associated with a reduced phase space.

The subalgebra $\sca_{\tL}=\Phi\times\cyl_{\tL}$ introduced by definition \ref{ics} corresponds to the reduced phase space $\tl{\cal E}\times\A_{\tL}$ which is {\em infinite} dimensional. The difference between this reduction and that applied by Kijowski is that here we do not reduce momentum degrees of freedom at all, while Kijowski reduces both momentum and configuration ones.

\subsubsection{The directed set $(\Lambda,\geq)$}

Now we are going to label the set of infinite classical systems by elements of directed sets. We define  
\[
(\Lambda,\geq)\ :=\ (\{\tL\},\geq),
\]
where the directing relation in the set $\{\tL\}$ of all tame subgroups of the hoop group $\h\G$ is defined as follows: $\tL'\geq\tL$ if $\tL$ is a subgroup\footnote{The existence of the directing relation on $\{\tL\}$ is guaranteed by the analyticity of the loops generating the hoop group $\h\G$ \cite{al-hoop}.} of $\tL'$ \cite{al-hoop}. Later on, as we will see in Subsection \ref{s-s-ISQ}, it will be necessary to restrict ourselves to a directed subset of this $\Lambda$. 

Thus we have finished Step 1 of the quantization procedure---the result of the step is the family of reduced classical systems labeled by the directed set.

\subsection{Step 2: Systems and subsystems}

The present section describes the {\em crucial step} of the quantization procedure, namely the definition and analysis of the relation 'system-subsystem' among the classical systems $\{\sca_\la\}_{\la\in\Lambda}$---once this step is done the remaining steps of the procedure will be (modulo some technical difficulties) rather straightforward. Since a step like this does not occur in the standard quantization procedure leading to the LQG model we advise the reader to study this section carefully.

\subsubsection{Preliminaries \label{pre}}

Consider {\em infinite} classical systems $\sca_{\la'}$ and $\sca_\la$  such that $\la'=\tL'\geq \la=\tL$. The essential observation is that the space $\cyl_{\tL'}$ describes more configuration degrees of freedom than $\cyl_{\tL}$ including those in $\cyl_{\tL}$. Indeed, the spaces are functions on, respectively, $\A_{\tL'}$ and $\A_{\tL}$ and
\[
\A_{\tL}=\pr_{\tL\tL'}(\A_{\tL'})
\] 
where $\pr_{\tL\tL'}$ is a linear\footnote{The linearity of the projection follows from \eqref{h-lin}.} projection from $\A_{\tL'}$ onto $\A_{\tL}$  such that 
\begin{equation}
\pr_{\tL}=\pr_{\tL\tL'}\circ\pr_{\tL'}.
\label{pr-LL}
\end{equation}
By virtue of Definition \ref{cyl-df} and \eqref{pr-LL} we can write
\[
\cyl_{\tL}\ni\Psi=\pr_{\tL}^*\,{\cal I}_{\cal L}^*\,\psi =\pr_{\tL'}^*\,\pr_{\tL\tL'}^*\,{\cal I}_{\cal L}^*\,\psi=\pr_{\tL'}^*\,{\cal I}_{\cal L'}^*\ [{\cal I}^{-1*}_{\cal L'}\pr_{\tL\tL'}^*{\cal I}_{\cal L}^*]\ \psi.
\]
Note now that ${\cal I}_{\cal L}\circ\pr_{\tL\tL'}\circ{\cal I}^{-1}_{\cal L'}$ is nothing else but a linear projections from $\R^{N'}\cong\A_{\tL'}$ onto $\R^{N}\cong\A_{\tL}$. Since the pull-back defined by the projection maps polynomials on $\R^N$ to polynomials on $\R^{N'}$ the function $[{\cal I}^{-1*}_{\cal L'}\pr_{\tL\tL'}^*{\cal I}_{\cal L}^*]\ \psi$ is bounded by polynomials in the sense of Definition \ref{mult-df}. Thereby the function is a multiplier of $\mathbb{S}_{N'}$, hence $\Psi\in\cyl_{\tL'}$ and 
\begin{equation}
\cyl_{\tL}\subset \cyl_{\tL'} 
\label{cyl-cyl}.
\end{equation}

In this way we see that 
\begin{equation}
\sca_{\la}\subset\sca_{\la'}.
\label{l-sub-l'}
\end{equation}

\begin{df}
We say that a classical system $\sca_\la$ is a subsystem of a classical system $\sca_{\la'}$ if and only if $\la'\geq\la$.   
\end{df}

However, in order to proceed with the quantization procedure the relation \eqref{l-sub-l'} has to be analyzed more carefully---this analysis will turn out to be necessary to define the projection $\pi_{\la\la'}$ from $\D_{\la'}$ onto $\D_{\la}$ required by the procedure. As it was already said the space $\D_{\la'}$ ($\D_{\la}$) of mixed states will be represented by the set of density matrices on the Hilbert space $\h_{\la'}$ $(\h_\la)$. Therefore $\pi_{\la\la'}$ will be defined according to the well known formula describing a projection from quantum system on onto its subsystem: given  $\la'\geq\la$, we will split $\h_{\la'}$ into a tensor product $\tl{\h}_{\la'\la}\ot\h_{\la'\la}$ where $\h_{\la'\la}$ will correspond to $\h_\la$ via a natural unitary map. Next, given $\rho_{\la'}\in\D_{\la'}$ we will evaluate the {\em partial trace} with respect to $\tl{\h}_{\la'\la}$ obtaining a density matrix on $\h_{\la'\la}$ which will be {\em identified} by the unitary map with a density matrix on $\h_\la$. Thus the question 'how to construct the projection $\pi_{\la\la'}$?' can be reduced to 'how to decompose $\h_{\la'}$ into $\tl{\h}_{\la'\la}\ot\h_{\la'\la}$?'. 

Since we are going to define the Hilbert space $\h_{\la'}$ in a natural way as $L^2(\A_{\tL'},d\mu_{\tL'})$, where $d\mu_{\tL'}$ is a Haar measure\footnote{Note that since $\A_{\tL'}$ is a linear space and hence a Lie group the notion of a Haar measure on $\A_{\tL'}$ is well defined.} on $\A_{\tL'}$ the decomposition $\tl{\h}_{\la'\la}\ot\h_{\la'\la}$ can be obtained from a decomposition of $\A_{\tL'}$. Taking into account the fact that $\pi_{\la\la'}$ is a quantum counterpart of the projection $\pr_{\tL\tL'}$  the decomposition should be of the form
\begin{equation}
\A_{\tL'}=\ker\pr_{\tL\tL'}\oplus\A_{\la'\la}.
\label{a-dec}
\end{equation}
Now we can set
\[
\tl{\h}_{\la'\la}=L^2(\ker \pr_{\tL\tL'},d\tl{\mu}_{\la'\la}) \ \ \ \text{and} \ \ \ \h_{\la'\la}=L^2(\A_{\la'\la},d{\mu}_{\la'\la})\cong\h_\la,
\]    
where $(i)$ $d{\mu}_{\la'\la}$ is a Haar measure on the linear space $\A_{\la'\la}$ corresponding via the projection $\pr_{\tL\tL'}$  to the Haar measure $d\mu_{\tL}$ on $\A_{\tL}$ and $(ii)$ the product $d\tl{\mu}_{\la'\la}\times d{\mu}_{\la'\la}$ coincides with $d\mu_{\tL'}$. 

We conclude then that the definition $\pi_{\la\la'}$ should be based on the decomposition \eqref{a-dec}. However, there are a lot of such decompositions which differ from each other by the choice of $\A_{\la'\la}$ and, as it is shown in Appendix \ref{dep-app}, the projection $\pi_{\la\la'}$  depends {\em essentially} on the choice of $\A_{\la'\la}$. Thus our task is to find a criterion which will allow us to distinguish $\A_{\la'\la}$ in a natural way. 

Note that a choice of $\A_{\la'\la}$ appearing in \eqref{a-dec} is equivalent to a choice of a linear embedding $\theta_{\la'\la}:\A_{\tL}\rightarrow \A_{\tL'}$ such that 
\begin{equation}
\pr_{\tL\tL'}\circ\theta_{\la'\la}=\id \ \ \text{and} \ \ \A_{\la'\la}=\im\theta_{\la'\la}.
\label{pr-th-id}
\end{equation}
Indeed, \eqref{a-dec} implies that $\pr_{\tL\tL'}$ restricted to $\A_{\la'\la}$ is a linear isomorphism onto $\A_{\tL}$. Then $\theta_{\la'\la}:=(\pr_{\tL\tL'}\big|_{\A_{\la'\la}})^{-1}$. On the other hand, if $\theta_{\la'\la}$ satisfies the first condition of \eqref{pr-th-id} then $\A_{\la'\la}$ given by the second one satisfies \eqref{a-dec}. Thus the equivalence follows. In fact, in the sequel we will use rather the embedding instead of $\A_{\la'\la}$ since the former will turn out to be more convenient. 

To finish the preliminary considerations let us emphasize that the choice of the subspace $\A_{\la'\la}$ and the embedding $\theta_{\la'\la}$  can be done by referring merely to the structure  of the {\em classical} systems $\sca_{\la'}$ and $\sca_\la$. This is why we decided to place this construction before the systems have been quantized.   

\subsubsection{Systems and subsystems---detailed consideration\label{s-s-ISQ}}

Assume that $\la'=\tL'\geq\la=\tL$ and consider the corresponding classical systems $\sca_{\la'}\supset\sca_{\la}$. The only difference between the systems comes from the difference between the groups $\tL'$ and $\tL$. Therefore while  defining $\A_{\la'\la}$ we should refer to the groups. 

Suppose then that the sets $\{l'_1,\ldots,l'_{N'}\}$ and $\{l_1,\ldots,l_N\}$ $(N'\geq N)$ of independent loops generate, respectively, the group $\tL'$ and $\tL$. Assume moreover that for $I=1,\ldots,N$ 
\begin{equation}
\tll'_I=\tll_I.
\label{h'h}
\end{equation}
Then we could define $\A_{\la'\la}$ as follows
\begin{df}
\begin{equation}
\A_{\la'\la}:=\{ \ [A]'\in\A_{\tL'}\ | \ h_{l'_{J'}}(A)=0 \ \ \text{for $J'=N+1,\ldots,N'$} \ \},
\label{im-theta}
\end{equation}
where $[A]'$ is the equivalence class of $A$ with respect to the relation $\sim_{\tL'}$ defining $\A_{\tL'}$.
\label{All-df-i}
\end{df} 
 However, the space $A_{\la'\la}$ given by the  definition is not unique because even if the condition \eqref{h'h} is satisfied the choice of the remaining loops $\{l'_{N+1},\ldots,l'_{N'}\}$ is still ambiguous---in general there are many choices of such loops and each of them may provide us with different measuring instrument  $\{h_{l'_{N+1}},\ldots,h_{l'_{N'}}\}$ (in other words, the choices of the loops are not holonomically equivalent in general). Therefore to remove the ambiguity we will restrict ourselves to tame groups generated by a special kind of piecewise analytical loops which will be called {\em almost analytic} loops. 

\begin{df}
We say that a loop $l$ (based at $y\in\Sigma$) is almost analytic if and only if
\begin{equation}
l=e^{-1}\circ\bar{l}\circ e,
\label{almost}
\end{equation}
where $(i)$ $\bar{l}$ is an analytic loop based at $y'\in\Sigma$ such that almost all points of the path defined by $\bar{l}$  are traced precisely once\footnote{This means that the loop can have self-intersections, but it cannot trace its own path more than once.}  and $(ii)$ $e$ is a piecewise analytic oriented path of the source at $y$ and the target at $y'$ ($e^{-1}$  denotes here the path obtained from $e$ by the change of the orientation). 
\label{almost-df}
\end{df}

Note that the only role of the path $e$ is to ensure that the base point of the loops $l$ is $y$---the path does not effect the value of the holonomy along $l$ i.e.    
\[
h_l=h_{\bar{l}}.
\]
Therefore we can modify Definition \ref{ind-loops} describing the notion of a set of independent loops in a way which will be suitable for further considerations. Using the notation introduced by Definition \ref{almost-df} we formulate 

\begin{df}
Suppose that a finite subset ${\cal L}=\{l_1,\ldots,l_M\}$ of $\mathbb{L}_y$ consists of almost analytic loops. ${\cal L}$ is called a set of independent loops if and only if  for each $I=1,\ldots,N$ the set 
\begin{equation}
\bar{l}_I\cap(\bar{l}_1\cup\ldots\cup\bar{l}_{I-1}\cup\bar{l}_{I+1}\cup\ldots\cup\bar{l}_N)
\label{l-lll}
\end{equation}
consists of a finite number of points.
\end{df}
It follows that for a set ${\cal L}=\{l_1,\ldots,l_M\}$ of independent almost analytic loops 
\[
\tll_I\neq\tll_J^{\pm 1}
\] 
unless $I\neq J$. 

We have the following lemma
\begin{lm}
\begin{enumerate}
\item Suppose $\{l_1,\ldots,l_N\}$ is a set of almost analytic loops such that for every $I\neq J$
\[
\tll_I\neq\tll_J^{\pm 1}.
\]  
Then $\{l_1,\ldots,l_N\}$ is a set of independent almost analytic loops. 
\item Assume that the tame groups $\tL'$ and $\tL$ are generated, respectively, by the sets ${\cal L}'=\{l'_1,\ldots,l'_{N'}\}$ and ${\cal L}=\{l_1,\ldots,l_N\}$ of almost analytic independent loops. If $\tL$ is a subgroup of $\tL'$ then there exist a unique map $\sigma:\{1,\ldots,N\}\mapsto\{1,\ldots,N'\}$ such that
\[
\tll_{I}=(\tll'_{\sigma(I)})^{\pm 1}.
\]     
The map $\sigma$ is injective.
\end{enumerate}

\label{almost+-}
\end{lm}

\noindent{\it Proof.} Statement 1. Given $I\neq J$, consider analytic loops $\bar{l}_I$ and $\bar{l}_J$ defining the loops $l_I$ and $l_J$ respectively. Assume that the set of intersections between $\bar{l}_I$ and $\bar{l}_J$ is infinite. Then analyticity of the loops implies immediately that $\bar{l}_I=\bar{l}_J^{\pm 1}$ and consequently $\tll_I=\tll_J^{\pm 1}$ which contradicts the assumptions of the lemma. Thus $\bar{l}_I\cap \bar{l}_J$ consists  of at most finite number of points and thereby the set \eqref{l-lll} is finite.

Statement 2. It is clear that the set ${\cal L}'$ satisfies the assumptions of Statement 1 of the lemma. Suppose that, given $I\in\{1,\ldots,N\}$, ${\cal L}'\cup l_I$ also satisfies the assumptions. Then it is a set of independent loops and the contradiction $\tll_I\not\in\tL'$ follows. Thus there has to exist $J'\in\{1,\ldots,N'\}$ such that $\tll_{I}=(\tll'_{J'})^{\pm 1}$. Moreover, this $J'$ has to be unique otherwise ${\cal L}'$ would not be a set of independent loops. This proves the existence and uniqueness of $\sigma$. The map has to be injective also, otherwise the set $\cal L$ would not be a set of independent loops. $\blacksquare$
          
Statement 1 of the above lemma means that within the domain of almost analytic loops the measuring instruments $\{h_l\}$ are {\em independent} modulo the trivial dependence $h_l=\pm h_{l'}$. More precisely, if $\{l_1,\ldots,l_N\}$ are almost analytic loops satisfying the assumption of the statement then the only function $H:\R^N\rightarrow\R$ which satisfies 
\[
H(h_{l_1},\ldots,h_{l_N})=0
\]
is $H=0$. 

Statement 2 of Lemma \eqref{almost+-} implies the following conclusions concerning the tame hoop groups $\tL',\tL$ appearing in the statement:
\begin{enumerate}
\item $\cal L$ is the {\em unique} (modulo the change of orientation and holonomical equivalence of the loops)  set of independent almost analytic loops generating $\tL$. 
\item if $\tL$ is a subgroup of $\tL'$ then there is a unique split of ${\cal L}'$ into two disjoint subsets such that the loops in one of them coincide (modulo the change of orientation and holonomical equivalence) with those in $\cal L$.
\end{enumerate}

The last conclusion guarantees that by restricting ourselves to almost analytic loops the ambiguity in Definition \eqref{All-df-i} of $\A_{\la'\la}$ is removed. Therefore since now we will consider only those infinite classical systems which are defined by tame hoop groups generated by sets of {\em almost analytic} independent loops. It is straightforward to check that these tame hoop groups also form a directed set with the directing relation defined by inclusion as before. This set will be still denoted by $(\Lambda,\geq)$.

Now let us turn back to the groups $\tL',\tL$ considered above. Without loss of generality we can assume that
\[
\tll_I=\tll'_I, \ \ \ I=1,\ldots,N.
\]
Let $(x'_{J'})$ and $(x_J)$ be linear coordinates on, respectively, $\A_{\tL'}$ and $\A_{\tL}$ given by the maps ${\cal I}_{{\cal L}'}$ and ${\cal I}_{\cal L}$ (see Equation \eqref{bij}). Then the map $\theta_{\la'\la}$ corresponding to $\A_{\la'\la}$ can be written as follows  
\[
\theta_{\la'\la}(x_1,\ldots,x_N):=(x_1,\ldots,x_N,0,\ldots,0),
\]
where the r.h.s. of the equation is expressed by means of the coordinates $(x'_{J'})$. 
 
Now it is easy to show that there holds the following important:
\begin{thr}
Given $\la''\geq\la\geq\la$
\begin{equation}
\theta_{\la''\la}=\theta_{\la''\la'}\circ\theta_{\la'\la}.
\label{theta-cons}
\end{equation}
\label{theta-cons-thr}
\end{thr}

\subsection{Step 3: Quantization of classical systems \label{step-3}}

In the previous step we gave a precise description of the relation 'system-subsystem' between the members of the family $\{\sca_\la\}_{\la\in\Lambda}$. Now we are going to focus on a fixed classical system ${\sca}_\la$ and {\em quantize} the system according to the standard rules of canonical quantization. Thus we will obtain the quantum system $\S_\la$ considered as an approximation of the quantum 'toy theory'. 

Let us extend the Lie algebra $\sca_\la$ which defines the classical system to an operator $*$-algebra $\hat{\sca}_{\la}$ generated by the following linear operators acting on the algebra $\cyl^S_{\tL}$ of Schwarz cylindrical functions compatible with the tame group $\tL$:
\begin{equation}
\begin{aligned}
&\cyl^S_{\tL}\,\ni\ \Psi'\mapsto\hat{\Psi}\Psi'\ :=\ \Psi\Psi', \ \ \ \Psi\in\cyl_{\tL},\\
&\cyl^S_{\tL}\,\ni\ \Psi'\mapsto \hat{\phi}_{S,f,\la}\Psi':=i\phi_{S,f}\Psi', \ \ \ \phi_{S,f}\in F.
\end{aligned}  
\label{repr}
\end{equation}
Clearly, $\Psi\Psi'$ is a function in $\cyl^S_{\tL}$, and $\hat{\phi}_{S,f,\la}\Psi'$  also belong to the space, which is guaranteed by Equation \eqref{e-psi-braket} and the fact that every derivative maps the algebra of Schwarz functions into itself. The $*$ involution on $\hat{\sca}_\la$ is defined as follows
\[
\hat{\Psi}^*:=\hat{\overline{\Psi}}, \ \ \ \hat{\phi}_{S,f,\la}^*:=\hat{\phi}_{S,f,\la}
\] 
(the correctness of the definition can be proved by a method described in \cite{ol}).

The next step of the quantization procedure is constructing a $*$-rep\-re\-sen\-ta\-tion of $\hat{\sca}_{\la}$ on a Hilbert space $\h_\la$. Let us begin this construction by defining the space. Let $(x^1,\ldots,x^N)$ be the canonical coordinates on $\R^N$ and let $dx^N$ be the Lebesgue measure on $\R^N$ which, obviously, coincides with one of the Haar measures on $\R^N$ thought as a Lie group. Given a set $\cal L$ of independent loops generating the tame group $\tL$ we define a measure $d\mu_{\tL}$ on $\A_{\tL}$ by the formula
\begin{equation}
\int_{\A_{\tL}}\psi\, d\mu_{\tL}:=\int_{\R^N} ({\cal I}^{-1*}_{\cal L} \psi)\,dx^N,
\label{dmuL}
\end{equation}
where ${\cal I}_{\tL}$ is given by \eqref{bij}. As it was shown in \cite{oko} the measure $d\mu_{\tL}$ does not depend on the choice of the set $\cal L$. Hence
\begin{equation}
\h_\la:=L^2(\A_{\tL},d\mu_{\tL}).
\label{h-lambda-df}
\end{equation}
is defined unambiguously.

Let us consider the following pull-back
\begin{equation}
C^{\infty}(\A_{\tL})\ni\Psi\mapsto\pr_{\tL}^*\Psi.
\label{pull}
\end{equation}
The map $\pr_{\tL}$ is a surjection, hence the pull-back is an injective map and $\pr^{*-1}_{\tL}$ is a well defined map from $\im \pr_{\tL}^*$ onto the space $C^{\infty}(\A_{\tL})$ of smooth functions on $\A_{\tL}$. Taking advantage of Definition  \ref{cyl-df} one can easily check that $\pr^{*-1}_{\tL}(\cyl^S_{\tL})$ is a dense subset of $\h_\la$. Equations \eqref{repr} provide us with a natural representation of $\hat{\sca}_\la$ on $\cyl^S_{\tL}$ which can be mapped by means of the map $\pr_{\tL}^*$ onto a $*$-representation $T_\la$ of $\hat{\sca}_\la$ on the Hilbert space $\h_\la$:
\begin{equation}
\hat{\sca}_\la\in\hat{a}\mapsto T_\la(\hat{a}):=\pr_{\tL}^{*-1}\circ \hat{a}\circ\pr_{\tL}^*.
\label{T-repr}
\end{equation}

Thus we have obtained the desired quantum system $\S_\la$ consisting of
\begin{enumerate}
\item {\em pure states} given by vectors in the Hilbert space $\h_\la$.
\item {\em mixed states} represented by the set $\D_\la$ of all density matrices on $\h_\la$.
\item the representation $T_\la$ of the $*$-algebra $\hat{\sca}_\la$ of observables on $\h_\la$.
\end{enumerate}

Note that some elements of $\hat{\sca}_\la$ (like e.g. the 'flux-like' operators $\{\hat{\phi}_{S,f,\la}\}$) are mapped by the representation $T_\la$ to unbounded operators on $\h_\la$, which cannot be evaluated on some elements of $\D_\la$. On the other hand every element of $\D_\la$ can be evaluated on the $C^*$-algebra ${\cal B}_\la$ of bounded operators on $\h_\la$. Therefore in the sequel we will use the algebras $\{{\cal B}_\la\}$ rather than $\{T_\la(\hat{\sca}_\la)\}$ to build the algebra of quantum observables for the 'toy theory'.         

\subsection{Step 4: The space of states for the 'toy theory'}

In the previous Subsection we have obtained the family $\{\S_\la\}$ of quantum systems labeled by the directed set $(\Lambda,\geq)$. Now we are going to build from these systems a large quantum system $\S$ which will consist of 
\begin{enumerate}
\item the space $\D$ of quantum states,
\item an algebra ${\cal B}$ of observables which can be evaluated on the states in $\D$.
\end{enumerate}
The system $\S$ can be considered as the {\em quantum 'toy theory'} with the Gauss constraint solved and the diffeomorphism one unsolved.

\subsubsection{Construction of the space $\D$ of quantum states} 

As mentioned earlier we are going to equip the set $\{\D_\la\}$ with a structure of a projective family---this means that we will have to define, for every pair $\la'\geq\la$, the projection $\pi_{\la\la'}:\D_{\la'}\rightarrow\D_\la$ (the 'forgetting' operator) in such a way that for every triple $\la''\geq\la'\geq\la$
\begin{equation}
\pi_{\la\la''}=\pi_{\la\la'}\circ\pi_{\la'\la''}.
\label{cons-0}
\end{equation}
The projections will be constructed by the method described in Subsection \ref{pre}. 

Assume then that $\la'\geq\la$. This means either $\la'=\la$ or $\la'>\la$ (i.e. $\la'\geq\la$ and $\la'\neq\la$). Let us begin the construction of the projections $\{\pi_{\la\la'}\}$ by considering the case $\la'>\la$. 
  
\begin{lm}
Suppose that $\la'>\la$. Then 
\[
\dim \A_{\tL'}>\dim \A_{\tL}.
\]    
\label{N'>N}
\end{lm} 

This lemma (proven in Appendix \ref{app-proof}) means that, given $\la'>\la$, the passage from the classical system ${\sca}_{\la'}$ to ${\sca}_\la$ involves a {\em non-trivial} reduction of the configuration degrees of freedom---the reduction is described by the projection $\pr_{\tL\tL'}$. Note now that the decomposition \eqref{a-dec} can be rewritten in the following form
\begin{equation}
\begin{array}{ccccc}
\A_{\tL'}&=&\ker \pr_{\tL\tL'}&\oplus&\im\theta_{\la'\la}\medskip\\
&&&&\Big\uparrow\vcenter{\rlap{$\scriptstyle{\theta_{\la'\la}}$}}\medskip\\
&&&&\A_{\tL}
\end{array}\ \ \,,
\label{AA-dec}
\end{equation}
where $\im\theta_{\la'\la}=\A_{\la'\la}$ and, by virtue of Lemma \ref{N'>N}, 
\begin{equation}
\ker\pr_{\tL\tL'}\neq 0.
\label{ker-not-0}
\end{equation}
Clearly, $\im\theta_{\la'\la}$ describes the degrees of freedom which do not undergo to the reduction and are identified  by means of $\theta^{-1}_{\la'\la}$ with the degrees of freedom of $\A_{\tL}$. Now we are going to use \eqref{AA-dec} to decompose the Hilbert space $\h_{\la'}=L^2(\A_{\tL'},d\mu_{\tL'})$ into a tensor product of two Hilbert spaces.   

To achieve the goal let us equip every space occurring in \eqref{AA-dec} with an appropriate measure. Of course, the spaces $\A_{\tL'}$ and $\A_{\tL}$ are already equipped with measures $d\mu_{\tL'}$ and $d\mu_{\tL}$ respectively which are given by Equation \eqref{dmuL}. Let us define a Haar measure $d\mu_{\la'\la}$ on $\im\theta_{\la'\la}$ by the push-forward
\begin{equation}
d\mu_{\la'\la}:=\theta_{\la'\la*}d\mu_{\tL}.
\label{mu-ll}
\end{equation}
It is easy to check that there exists a unique Haar measure $d\tl{\mu}_{\la'\la}$ on  $\ker\pr_{\tL\tL'}$ such that $d\mu_{\tL'}=d\tl{\mu}_{\la'\la}\times d\mu_{\la'\la}$. Thus we have obtained the following diagram  
\begin{equation}
\begin{array}{ccccc}
d\mu_{\tL'}&=&d\tl{\mu}_{\la'\la}&\times&d\mu_{\la'\la}\medskip\\
&&&&\Big\uparrow\vcenter{\rlap{$\scriptstyle{\theta_{\la'\la*}}$}}\medskip\\
&&&& d\mu_{\tL}
\end{array}\ \ \,,
\label{mmm}
\end{equation}
which corresponds to \eqref{AA-dec}. The measures just introduced allow us to define the following Hilbert spaces
\[
\tl{\h}_{\la'\la}:=L^2(\ker\pr_{\tL\tL'},d\tl{\mu}_{\la'\la})\ \ \text{and} \ \ \h_{\la'\la}:=L^2(\im\theta_{\la'\la},d\mu_{\la'\la})
\]
and a unitary map 
\begin{equation}
\h_{\la'\la}\ni\Psi\mapsto U_{\la'\la}\Psi:=\theta_{\la'\la}^{*}\Psi\in\h_\la,
\label{Ull}
\end{equation}
which will be used to identify the Hilbert spaces $\h_{\la'\la}$ and $\h_\la$.

Thus we obtained the desired decomposition:
\begin{equation}
\begin{array}{ccccc}
\h_{\la'}&=&\tl{\h}_{\la'\la}&\ot&\h_{\la'\la}\medskip\\
&&&&\Big\downarrow\vcenter{\rlap{$\scriptstyle{U_{\la'\la}=\,\theta^{*}_{\la'\la}}$}}\medskip\\
&&&& \h_\la
\end{array}\ \ \ \ \ \ \ \ \,.
\label{hhh}
\end{equation}
Comparing the above result with \eqref{AA-dec} we see that the Hilbert space $\tl{\h}_{\la'\la}$  corresponds to classical degrees of freedom undergoing to the reduction, while $\h_{\la'\la}$ does to classical degrees which survive the reduction. 

To define the projection $\pi_{\la\la'}$ we will also need an isomorphism $u_{\la'\la}$ from the algebra ${\cal B}_{\la'\la}$ of bounded operators on $\h_{\la'\la}$ onto the algebra ${\cal B}_\la$ of bounded operators on $\h_\la$ given by
\begin{equation}
{\cal B}_{\la'\la}\ni a \mapsto u_{\la'\la}a:=U_{\la'\la}\circ a \circ U^{-1}_{\la'\la}\in{\cal B}_\la.
\label{ull}
\end{equation}

Now, having the decomposition \eqref{hhh} and the above isomorphism we are ready to define the projection $\pi_{\la\la'}$. Let $\{\tl{\Psi}_n\}$ and $\{{\Psi}_n\}$ be orthonormal bases of, respectively, $\tl{\h}_{\la'\la}$ and ${\h}_{\la'\la}$. Then the partial trace
\[
\D_{\la'}\ni\rho\mapsto\tl{\tr}_{\la'\la}\,\rho:= \sum_{nm}[\ \sum_{k}\ \scal{\tl{\Psi}_k\ot{\Psi}_n}{\,\rho \, (\tl{\Psi}_k\ot{\Psi}_m)}\ ] \ \ket{\Psi_{n}}\,\bra{\Psi_{m}}
\]  
produces a density matrix on ${\h}_{\la'\la}$. In other words $\tl{\tr}_{\la'\la}$ reduces those degrees of freedom of the quantum system $\S_{\la'}$ which correspond to the classical degrees of freedom contained in $\ker\pr_{\tL\tL'}$. What remains to do is to map the resulting density matrix into one in $\D_\la$. Thus we have arrived at     

\begin{df}
Given $\la'\geq\la$, the projection $\pi_{\la\la'}:\D_{\la'}\rightarrow\D_\la$ is defined by the following formula:
\[
\pi_{\la\la'}:=
\begin{cases}
u_{\la'\la}\circ\tl{\tr}_{\la'\la} & \text{if $\la'>\la$}\\
\id & \text{if $\la'=\la$}
\end{cases}.
\]
\label{forget}
\end{df}

Before we prove that the projections just defined  satisfy the consistency condition \eqref{cons-0} let us make some preparatory technical considerations concerning the case $\la''>\la'>\la$. 

Let us note first that the decomposition \eqref{hhh} applied to the Hilbert spaces $\h_{\la''}$, $\h_{\la'}$ and $\h_\la$ gives us the following diagram:  
\begin{equation}
\begin{array}{cccccc}
\h_{\la''}=\tl{\h}_{\la''\la'}\,\ot&\!\!\!\!\h_{\la''\la'}&&=&\tl{\h}_{\la''\la}\,\ot&\!\!\!\!\h_{\la''\la}\medskip\\
&\!\!\!\!\Big\downarrow\vcenter{\rlap{$\scriptstyle{U_{\la''\la'}}$}}&&&&\!\!\!\!\Big|\medskip\\
&\!\!\!\!\h_{\la'}&=\tl{\h}_{\la'\la}\,\ot&\!\!\!\!\h_{\la'\la}&&\!\!\!\!\Big|\vcenter{\rlap{$\scriptstyle{U_{\la''\la}}$}}\medskip\\
&&&\!\!\!\!\Big\downarrow\vcenter{\rlap{$\scriptstyle{U_{\la'\la}}$}}&&\!\!\!\!\Big\downarrow\medskip\\
&&&&\!\!\!\!\h_\la&
\end{array}.
\label{diag}
\end{equation}
Now we are going to reconcile the two decompositions of $\h_{\la''}$ occurring in the top row of the diagram. 

Note first that by applying twice the decomposition \eqref{AA-dec} 
we get
\[
\A_{\tL''}=\ker\pr_{\tL'\tL''}\oplus\im\theta_{\la''\la'}\big|_{\ker\pr_{\tL\tL'}}\oplus\im\theta_{\la''\la'}\big|_{\im\theta_{\la'\la}}.
\]
Theorem \ref{theta-cons-thr} allow us to simplify the last term of the above formula. Thus we obtain
\begin{multline}
\A_{\tL''}=\ker\pr_{\tL'\tL''}\oplus\underset{\im \theta_{\la''\la'}}{\underbrace{\im\theta_{\la''\la'}\big|_{\ker\pr_{\tL\tL'}}\oplus\im\theta_{\la''\la}}}=\\=\underset{\ker\pr_{\tL\tL''}}{\underbrace{\ker\pr_{\tL'\tL''}\oplus\im\theta_{\la''\la'}\big|_{\ker\pr_{\tL\tL'}}}}\oplus\im\theta_{\la''\la}.
\label{A''-dec}
\end{multline}
Next, let us decompose the measure $d\mu_{\tL''}$ on $\A_{\tL''}$ in an analogous manner. We have from \eqref{mmm} 
\[
d\mu_{\tL''}=d\tl{\mu}_{\la''\la'}\times d{\mu}_{\la''\la'}
\]
By virtue of \eqref{mmm} and Theorem \ref{theta-cons-thr} we get 
\begin{multline}
d\mu_{\la''\la'}=\theta_{\la''\la'*}d\mu_{\tL'}=\theta_{\la''\la'*}(d\tl{\mu}_{\la'\la}\times d\mu_{\la'\la})=\\=(\theta_{\la''\la'}\big|_{\ker\pr_{\tL\tL'}})_*(d\tl{\mu}_{\la'\la})\times (\theta_{\la''\la'}\big|_{\im\theta_{\la'\la}})_* (d\mu_{\la'\la})
\label{mmm-1}
\end{multline}
where $(\theta_{\la''\la'}\big|_{\ker\pr_{\tL\tL'}})_*(d\tl{\mu}_{\la'\la})$ is a Haar measure on $\im(\theta_{\la''\la'}\big|_{\ker \pr_{\tL\tL'}})$. It follows from \eqref{mu-ll} that
\[
(\theta_{\la''\la'}\big|_{\im\theta_{\la'\la}})_* (d\mu_{\la'\la})=d\mu_{\la''\la}.
\]
In this way we obtain
\[
d\mu_{\tL''}=d\tl{\mu}_{\la''\la'}\times (\theta_{\la''\la'}\big|_{\ker\pr_{\tL\tL'}})_*(d\tl{\mu}_{\la'\la})\times d\mu_{\la''\la}.
\]
It is clear that the product of the two first measures on the r.h.s. of the above equation is a measure on $\ker\pr_{\tL\tL''}$, and Equation \eqref{mmm} allows us to conclude that the product coincides with $d\tl{\mu}_{\la''\la}$. The desired decomposition of $d\mu_{\tL''}$ reads
\begin{multline*}
d\mu_{\tL''}=d\tl{\mu}_{\la''\la'}\times \underset{d\mu_{\la''\la'}}{\underbrace{(\theta_{\la''\la'}\big|_{\ker\pr_{\tL\tL'}})_*(d\tl{\mu}_{\la'\la})\times d\mu_{\la''\la}}}=\\=
\underset{d\tl{\mu}_{\la''\la}}{\underbrace{d\tl{\mu}_{\la''\la'}\times (\theta_{\la''\la'}\big|_{\ker\pr_{\tL\tL'}})_*(d\tl{\mu}_{\la'\la})}}\times d\mu_{\la''\la}.
\end{multline*} 
The above result and the decomposition \eqref{A''-dec} give us immediately
\begin{equation}
\h_{\la''}=\tl{\h}_{\la''\la'}\ot\underset{\h_{\la''\la'}}{\underbrace{\tl{\h}_{\la''\la'\la}\ot\h_{\la''\la}}}=\underset{\tl{\h}_{\la''\la}}{\underbrace{\tl{\h}_{\la''\la'}\ot\tl{\h}_{\la''\la'\la}}}\ot\h_{\la''\la},
\label{hhhh}
\end{equation}
where
\[
\tl{\h}_{\la''\la'\la}=L^2(\,\im(\theta_{\la''\la'}\big|_{\ker \pr_{\tL\tL'}})\, , \,(\theta_{\la''\la'}\big|_{\ker\pr_{\tL\tL'}})_*(d\tl{\mu}_{\la'\la})\,). 
\]

It is easy to see that the following maps
\begin{align*}
&\tl{\h}_{\la''\la'\la}\ni\Psi\mapsto \tl{V}_{\la''\la'}\Psi=(\theta^{c}_{\la''\la'}\big|_{\ker\pr_{\tL\tL'}})^*\Psi\in\tl{\h}_{\la'\la},\\
&\h_{\la''\la}\ni\Psi\mapsto V_{\la''\la'}\Psi=(\theta^{c}_{\la''\la'}\big|_{\im\theta_{\la'\la}})^*\Psi\in\h_{\la'\la}.
\end{align*}
are unitary and satisfy
\begin{equation}
U_{\la''\la'}=\tl{V}_{\la''\la'}\ot V_{\la''\la'} \ \ \ \text{and} \ \ \ U_{\la'\la}\circ V_{\la''\la'}=U_{\la''\la}
\label{uvv}
\end{equation}
(the first equation follows from Equation \eqref{Ull}, while the last one does from Theorem \ref{theta-cons-thr}).

Thus we have finished the preparatory considerations and are ready to state and prove

\begin{thr}
For every triple $\la''\geq\la'\geq\la$ 
\begin{equation}
\pi_{\la\la''}=\pi_{\la\la'}\circ\pi_{\la'\la''}.
\label{cons}
\end{equation}
\end{thr}

\noindent{\it Proof.} Assume that $\la''>\la'>\la$ otherwise the proof is trivial. Consider the map $\tl{\tr}_{\la''\la}$ occurring in the definition of $\pi_{\la\la''}$. This map reduces the degrees of freedom described by the Hilbert space $\tl{\h}_{\la''\la}$, hence by virtue of \eqref{hhhh} it can be expressed as
\[
\tl{\tr}_{\la''\la}=\tl{\tr}_{0}\circ\tl{\tr}_{\la''\la'},
\]  
where $\tl{\tr}_{0}$ reduces the degrees of freedom described by $\tl{\h}_{\la''\la'\la}$. The first equation of \eqref{uvv} is equivalent to the following diagram
\[
\begin{array}{ccccc}
\h_{\la''\la'}&=&\tl{\h}_{\la''\la'\la}&\ot&\h_{\la''\la}\medskip\\
\Big\downarrow\vcenter{\rlap{$\scriptstyle{U_{\la''\la'}}$}}&&\Big\downarrow\vcenter{\rlap{$\scriptstyle{\tl{V}_{\la''\la'}}$}} &&\Big\downarrow\vcenter{\rlap{$\scriptstyle{V_{\la''\la'}}$}}\medskip\\
\h_{\la'}&=&\tl{\h}_{\la'\la}&\ot&\h_{\la'\la}
\end{array}\ \ \,.
\] 
Denote by $v_{\la''\la'}$ an isomorphism from the algebra of bounded operators on $\h_{\la''\la}$ onto one of bounded operators on $\h_{\la'\la}$ defined by the unitary map $V_{\la''\la'}$. It is clear now that 
\[
\tl{\tr}_{0}= v_{\la''\la'}^{-1}\circ\tl{\tr}_{\la'\la}\circ u_{\la''\la'};
\]    
therefore the projection $\pi_{\la\la''}$ can be expressed as follows
\begin{multline*}
\pi_{\la\la''}=u_{\la''\la}\circ\tl{\tr}_{\la''\la}=u_{\la''\la}\circ\tl{\tr}_{0}\circ\tl{\tr}_{\la''\la'}=(u_{\la''\la}\circ v^{-1}_{\la''\la'})\,\circ \,\tl{\tr}_{\la'\la}\,\circ\,(u_{\la''\la'}\circ\tr_{\la''\la'})=\\=(u_{\la''\la}\circ v^{-1}_{\la''\la'})\,\circ \,\tl{\tr}_{\la'\la}\,\circ\,\pi_{\la'\la''}.
\end{multline*}
The second equation of \eqref{uvv} implies that $u_{\la''\la}\circ v^{-1}_{\la''\la'}=u_{\la'\la}$, hence
\[
\pi_{\la\la''}=u_{\la'\la}\circ\tl{\tr}_{\la'\la}\circ\pi_{\la'\la''}=\pi_{\la\la'}\circ\pi_{\la'\la''}.
\]  
$\blacksquare$

Thus we have constructed the desired projective family $\{\D_\la,\pi_{\la\la'}\}$. This gives us the space $\D$ of states of the 'toy theory' as the projective limit of the family 
\begin{equation}
\D:=\underleftarrow{\lim} \,\D_\lambda.
\label{D-df}
\end{equation}

\subsubsection{Algebra of observables \label{alg-sub}}

Following the construction in \cite{kpt} we will define the algebra $\cal B$ of observables for the quantum 'toy theory'.

Let ${\cal B}_\la$ denote the $C^*$-algebra of bounded operators on the Hilbert space $\h_\la$. Given $a_\la\in{\cal B}_{\la}$ and $\rho_\la\in\D_\la$, there is a complex number associated with them:
\[
\langle a_\la, \rho_\la \rangle :=\tr(a_\la\rho_\la)\in\C. 
\]        

One can prove \cite{kpt} the following 
\begin{thr}
  ($i$) Given any $a_\la\in{\cal B}_\la$, there exists a unique operator $a_{\la'}\in{\cal B}_{\la'}$ such that for every $\rho_{\la'}\in\D_{\la'}$
\[
\langle a_{\la'},\rho_{\la'}\rangle =\langle a_\la,\pi_{\la\la'}\rho_{\la'}\rangle.
\]  
$(ii)$ The map
\[
{\cal B}_{\la}\ni a_\la \mapsto \pi^*_{\la'\la}a_\la:=a_{\la'}\in {\cal B}_{\la'}
\]
is a $C^*$-isomorphism of ${\cal B}_\la$ onto its image in ${\cal B}_{\la'}$.   

\noindent $(iii)$ for any triple $\la''\geq\la'\geq\la$ 
\[
\pi^*_{\la''\la}=\pi^*_{\la''\la'}\circ\pi^*_{\la'\la}.
\]  
\label{B-pi-thr}
\end{thr}

We conclude now that 
\[
\{\,{\cal B}_\la\, , \, \pi^*_{\la\la'}\,\}_{\la\in\Lambda}
\]
is an inductive family of (unital) $C^*$-algebras. Consequently, the inductive limit of the family,
\begin{equation}
{\cal B}:=\underrightarrow{\lim} \,{\cal B}_\la,
\label{B-alg}
\end{equation}
is a unital $C^*$-algebra also. It is easy to realize that every element of $\D$ defines a positive linear functional (a state) on $\cal B$ (note however that since $\D_\la$ does not coincide with the set of all states on ${\cal B}_\la$, the set $\D$ does not coincide with the set of all states on $\cal B$ either). We will consider the algebra $\cal B$ as an {\em algebra of (Yang-Mills gauge invariant) observables} for the quantum 'toy theory'.

\subsection{Step 5: Solutions of the diffeomorphism constraint \label{sol-diff}}

The goal of this section is to define an action of the {\em analytic diffeomorphisms} of $\Sigma$ on the space $\D$ of quantum states of the 'toy theory'  and find states preserved by the action---these states will be considered as quantum solutions of the diffeomorphism constraint \eqref{diff}.

Let us begin by describing the action of diffeomorphisms on the hoop group $\cal HG$. Given an analytic diffeomorphism $\chi$ and a loop $l\in\mathbb{L}_y$, the loop $\chi(l)$ is  based at the point $\chi(y)$. If $e$ is an (piecewise) analytic edge originating in $y$ and ending in $\chi(y)$ then for all our purposes we can identify $\chi(l)$ with the loop $e^{-1}\circ\chi(l)\circ e$ based at $y$ since
\begin{equation}
h_{\chi(l)}=h_{e^{-1}\circ\chi(l)\circ e}.
\label{phi-l}
\end{equation}
Moreover, the above equation holds independently of the choice of $e$. Thus the loop $e^{-1}\circ\chi(l)\circ e$ {\em defines} a unique hoop $\chi(\tll)\in{\cal HG}$ such that   
\[
h_{\chi(\tll)}=h_{\chi(l)}.
\]

In fact, although diffeomorphisms do not preserve the base point $y$ occurring in the definition of the hoop group $\cal HG$, the action 
\[
\tll\mapsto\chi(\tll)
\]
is an automorphism of the group \cite{oko}. This is why the quantization procedure does not depend of the choice of the point $y$ as it was mentioned earlier.    

Now we turn to the action of diffeomorphisms on the  measuring instruments $\varphi_{S,f}$ and $h_{\tll}$. The action \eqref{diff-ac} of an {\em analytic} diffeomorphism $\chi$ on the canonical variables induces the action of the diffeomorphism on the instruments
\[
({\tl{\chi}}\varphi_{S,f})(\tl{E}):=\varphi_{S,f}(\chi^{-1}_*\tl{E}) \ \ \ \text{and} \ \ \ ({\tl{\chi}}h_{\tll})(A):=h_{\tll}(\chi^{*}A).
\] 
Then it is easy to show that
\[
{\tl{\chi}}\varphi_{S,f}=\varphi_{\chi(S),\chi^{-1*}f} \ \ \ \text{and} \ \ \ {\tl{\chi}}h_{\tll}=h_{\chi(\tll)}.
\]
 
Using the above formulas we can easily define a {\em linear} action of diffeomorphism on the ACZ algebra $\sca$:
\begin{align}
&\cyl_{\tL}\,\ni\,\Psi=\pr^*_{\tL}\,{\cal I}_{\cal L}^*\,\psi\ \mapsto \ \tl{\chi}\Psi:=\pr^*_{\chi(\tL)}{\cal I}_{\chi({\cal L})}^*\,\psi\,\in\,\cyl_{\chi(\tL)},\label{diff-cyl}\smallskip\\
&\Phi\ni\phi_{S,f}\mapsto \tl{\chi}\phi_{S,f}:=\phi_{\chi(S),\chi^{-1*}f}\in\Phi.\nonumber
\end{align}
Since above we have used the maps ${\cal I}_{\cal L}, {\cal I}_{\chi({\cal L})}$\footnote{The loops $\chi({\cal L})=\{\chi(l_1),\ldots,\chi(l_n)\}$ are based at $\chi(y)$ but they still can be used by virtue of \eqref{phi-l}.} it is necessary to show that the definition does not depend on the choice of the independent loops ${\cal L}=\{l_1,\ldots,l_n\}$. Suppose then that ${\cal L}'=\{l'_1,\ldots,l'_n\}$ is another set of independent loops generating the tame group $\tL$. Then, as stated in Lemma \ref{A=R} the map $M:={\cal I}_{\cal L}\circ{\cal I}^{-1}_{{\cal L}'}$ is a linear automorphism of $\R^n$. As it is shown in \cite{oko} $M$ is fully determined by the decomposition of hoops $\{\tll_1,\ldots,\tll_n\}$ in terms of $\{\tll'_1,\ldots,\tll'_n\}$ which is of the same form as the decomposition of $\{\chi(\tll_1),\ldots,\chi(\tll_n)\}$ in terms of $\{\chi(\tll'_1),\ldots,\chi(\tll'_n)\}$. This means that $M$ is diffeomorphism invariant i.e. $M={\cal I}_{\chi({\cal L})}\circ{\cal I}^{-1}_{\chi({\cal L}')}$. Consequently, the definition under consideration is correct. It is obvious that           
\[
\tl{\chi}({\sca_\la})=\sca_{\tl{\chi}(\la)},
\]
where  $\tl{\chi}(\lambda):=\chi(\tL)$.

Let us now describe how the diffeomorphisms act on the space of states $\D$. Consider first the set of pure states of a system ${\sca}_{\la}$  which can be identified with the Hilbert space $\h_\la$. The action of diffeomorphism reads (compare with \eqref{diff-cyl}) 
\begin{equation}
\h_{\la}\ni\Psi={\cal I}^*_{\cal L}\psi\mapsto\tl{\chi}\Psi:={\cal I}^*_{\cal \chi({\cal L})}\psi\in\h_{\tl{\chi}(\la)},
\label{phi-uni}
\end{equation}
where $\psi\in L^2(\R^N,dx)$ with $N=\dim\A_{\tL}$. Thus we obtained  a unitary map from $\h_\la$ onto $\h_{\tl{\chi}(\la)}$ which defines an isomorphism 
\begin{equation}
\check{\chi}:{\cal B}_\la\rightarrow{\cal B}_{\tl{\chi}(\la)}.
\label{diff-la}
\end{equation}
Since $\D_\la$ can be naturally seen as a subset of ${\cal B}_\la$ the map $\check{\chi}$ defines an isomorphism from $\D_\la$ onto $\D_{\tl{\chi}(\la)}$ which will be denoted by the same symbol. 

\begin{lm}
For every analytic isomorphism $\chi$ 
\[
\check{\chi}\circ\pi_{\la\la'}\circ\check{\chi}^{-1}=\pi_{\tl{\chi}(\la)\tl{\chi}(\la')}.
\]
\end{lm}
    
\noindent {\it Proof.} Let $\la'=\tL'\geq\la=\tL$. Denote by ${\cal L}'$ $({\cal L})$ a set of independent loops generating $\tL'$ $(\tL)$. Then the projection and the set   
\begin{align*}
&\pr:={\cal I}_{\tl{\chi}({\cal L})}\circ \pr_{\tl{\chi}(\tL)\tl{\chi}(\tL')}\circ{\cal I}^{-1}_{\tl{\chi}({\cal L}')},\\ 
&\R_{\la\la'}:={\cal I}_{\tl{\chi}({\cal L}')}(\A_{\tl{\chi}(\la')\tl{\chi}(\la)})
\end{align*} 
do not depend on the choice of $\chi$. It is clear that
\[
\R^{N'}=\ker\pr\oplus\R_{\la'\la},
\]
where $N'$ denotes the number of loops in ${\cal L}'$. The above decomposition defines a projection $\pi$ from the set of density matrices on $L^2(\R^{N'},dx)$ onto the set of density matrices on $L^2(\R^{N},dx)$, where $N$ is the number of loops in ${\cal L}$.   

Denote by $\eta_{\cal L}$ an isomorphism from the algebra of bounded operators on $L^2(\R^{N},dx)$ onto ${\cal B}_\la$ induced by the unitary map ${\cal I}^*_{\cal L}:L^2(\R^N,dx)\rightarrow\h_\la$. We can conclude now that
\[
\pi=\eta^{-1}_{\tl{\chi}({\cal L})}\circ\pi_{\tl{\chi}(\la)\tl{\chi}(\la')}\circ\eta_{\tl{\chi}({\cal L}')}
\]     
independently on the choice of $\chi$. It means in particular that
\[
\eta^{-1}_{{\cal L}}\circ\pi_{\la\la'}\circ\eta_{{\cal L}'}=\eta^{-1}_{\tl{\chi}({\cal L})}\circ\pi_{\tl{\chi}(\la)\tl{\chi}(\la')}\circ\eta_{\tl{\chi}({\cal L}')}.
\] 
Now it is enough to note that $\check{\chi}=\eta_{\tl{\chi}({\cal L})}\circ\eta^{-1}_{\cal L}$. $\blacksquare$   

The lemma just proven  means that the following map 
\begin{equation}
\check{\chi}\rho=\check{\chi}\{\rho_\la\}:=\{\check{\chi}\rho_\la\}
\label{diff-state}
\end{equation}
maps $\D$ onto itself.

We say that a state $\rho\in\D$ is a solution of the diffeomorphism constraint if and only if it is diffeomorphism invariant, i.e. for every (analytic) diffeomorphism $\chi$   
\begin{equation}
\check{\chi}\rho=\rho.
\label{diff-sol}
\end{equation}

\subsubsection{Diffeomorphism invariant states in $\cal D$ \label{diff-inv-ISQ}}

Consider a Hilbert space $\h_\la$ with $\la=\tL$. We know that modulo the change of orientation and the holonomical equivalence the set $\{l_1,\ldots,l_N\}$ of almost analytic independent loops generating the tame group $\tL$ is {\em unique}. Thus there is a distinguished isomorphism 
\begin{equation}
\h_\la\cong\h_{\tL(\tll_1)}\ot\ldots\ot\h_{\tL(\tll_N)},
\label{h-la}
\end{equation}
where $\tL(\tll_I)$ is a tame group of hoops generated by the hoop $\tll_I$. Clearly, this isomorphism is generated by the maps $\{\theta_{\la\la_I}\}$, where $\la_I=\tL(\tll_I)$.

As we know already, $\la'=\tL'\geq\la=\tL$ if and only if the set  $\{l_1,\ldots,l_N\}$ of independent loops generating $\tL$ is (again modulo the change of orientation and the holonomical equivalence) a subset of the set $\{l'_1,\ldots,l'_{N'}\}$ of independent loops generating $\tL'$. Consider now the projection $\pi_{\la\la'}$ which reduces some quantum degrees of freedom described by the Hilbert space
\[
\h_{\la'}\cong\h_{\tL(\tll'_1)}\ot\ldots\ot\h_{\tL(\tll'_{N'})}.
\]
It is easy too see that degrees of freedom undergoing the reduction are described by those Hilbert spaces $\{\h_{\tL(\tll'_I)}\}$ for which the corresponding loops $\{l'_I\}$ do not belong (modulo the change of orientation and the holonomical equivalence)  to $\{l_1,\ldots,l_N\}$.    

Thus every Hilbert space $\h_\la$ can be naturally and unambiguously decomposed into the tensor product \eqref{h-la}. Moreover, the decomposition provides us with a simple description of the projections $\pi_{\la\la'}$. 

Let us start the construction of diffeomorphism invariant states by defining an equivalence relation on the set $\mathbb{AL}_y$ of all almost analytic loops based at the fixed point $y$. We say that the loops $l_1$ and $l_2$ are equivalent if and only if there exist an analytic diffeomorphism $\chi$ on $\Sigma$ such that
\begin{equation}
\chi(\tL(\tll_1))=(\tL(\tll_2)).
\label{loop-diff}
\end{equation}
      
Now, denoting by $[l]$ the equivalence class of an almost analytic loop $l$, consider the following map 
\[
[l]\mapsto\rho([l]),
\]
where $\rho([l])$ is a density matrix on the Hilbert space $L^2(\R,dx)$ such that it is preserved by the unitary map on the Hilbert space defined by $\R\ni x\mapsto-x$. This density matrix can be mapped naturally into a density matrix $\rho(\tL(\tll))$ on $\h_{\tL(\tll)}$. Indeed, the Hilbert space $\h_{\tL(\tll)}$ is obviously isomorphic to $L^2(\R,dx)$. In fact, there exist two distinguished isomorphisms between the Hilbert spaces: one given by ${\cal I}_{\{l\}}$ and another one given by ${\cal I}_{\{l^{-1}\}}$.  Then it is easy to see that thanks to the assumed special property of $\rho([l])$ the two distinguished isomorphisms between $L^2(\R,dx)$ and $\h_{\tL(\tll)}$ map $\rho([l])$ to the same density matrix. Thus we have constructed  a map
\[
\tL(\tll)\mapsto\rho(\tL(\tll))\in\D_{\tL({\tll})}
\]
which commutes with the action of every diffeomorphism $\chi$:
\begin{equation}
\check{\chi}\,\rho(\tL(\tll))=\rho(\,\chi(\tL(\tll))\,).
\label{comm-diff}
\end{equation}

To construct a diffeomorphism invariant state it is enough to  decompose every Hilbert space $\h_\la$ as it was done in Equation \eqref{h-la} and define
\[
\rho_\la:=\rho(\tL(\tll_1))\ot\ldots\ot{\rho}(\tL(\tll_N))\ \in \ \D_{\tL(\tll_1)}\ot\ldots\ot\D_{\tL(\tll_N)}\subset\D_\la.
\]   
It is straightforward to check that for every pair $\la'\geq\la$ 
\[
\pi_{\la\la'} \,\rho_{\la'}=\rho_{\la}.
\]
Thus $\rho=\{\rho_\la\}$ is an element of $\D$. Consider now the action of an analytic diffeomorphism $\chi$ on the density matrix $\rho_\la$. Equation \eqref{comm-diff} gives us immediately  
\[
\check{\chi}\rho_\la=\rho_{\tl{\chi}(\la)}
\]  
which means that the state $\rho$ is {\em diffeomorphism invariant}. 

Of course, every convex combination of (a finite number of) diffeomorphism invariant states is a diffeomorphism invariant state also. Thus we have obtained a plenteous set $\D_{\rm ph}$ of diffeomorphism invariant states which are, of course,  also Yang-Mills gauge invariant. Therefore we can treat the set $\D_{\rm ph}$ as the set of {\em physical} states of the 'toy theory'. Moreover, since $\D_{\rm ph}\subset\D$ Yang-Mills gauge invariant quantum observables in $\cal B$ can be evaluated on the states in $\D_{\rm ph}$.

This completes the quantization of the 'toy theory': the resulting quantum model ${\cal S}_{\rm ph}$ is the $C^*$-algebra $\cal B$ constructed in Subsection \ref{alg-sub} equipped with the set $\D_{\rm ph}$ of physical states.   

\section{Discussion}

In this paper we have quantized a diffeomorphism invariant theory of connection with a non-compact structure group using a strategy based on the projective techniques by Kijowski combined with the LQG methods. It should be stated clearly that our goal was not the quantum model of the 'toy theory' by itself---we rather wanted to check whether the strategy can be applied to quantization of a theory of connections. Although we were interested in the application to the non-compact case it seems that the strategy does not distinguish between the compact and non-compact case since the 'forgetting' operator \eqref{forget} can be defined once a split of a (separable) Hilbert space into the tensor product of two Hilbert spaces is given---it is absolutely irrelevant whether the Hilbert spaces are defined as $L^2$ spaces over compact or non-compact spaces. 

However, the fact that the strategy passed the test cannot be considered as a success because we do not know whether it was because of the power of the strategy or because of simplicity of the 'toy theory'. In other words, we still do not know whether the strategy can be applied to any non-trivial theory, in particular, to GR in terms of the complex\footnote{Of course, in this case we would also have to solve the problem of the reality conditions. Since we do not have any idea how to do this we will neglect the problem in the present discussion.} Ashtekar or the real Ashtekar-Barbero variables. Such an application does not seem to be easy: when one tries to apply the strategy to GR he immediately encounters problems caused by {\em non-commutativity} of the structure groups, respectively, $SL(2,\C)$ or $SU(2)$ e.g. the Yang-Mills gauge transformations are much more complicated than in the commutative case, the scalar constraint does not vanish. Thus it is not obvious that the strategy can be successfully applied in the case of any {\em non-commutative} structure group. It is  possible that the strategy just exchanges the non-compactness problem into the non-commutativity one i.e. that the inductive techniques work well only in the compact case, while the projective ones do well only in the commutative one.

\subsection{Almost analytic loops}

The first question one can ask is whether the application of the almost analytic loops (Definition \ref{almost-df}) can be justified, i.e. whether these loops provide sufficiently large set of functions $\{h_{l}\}$. To answer the question it is enough to check whether the functions separate points on ${\cal A}/{\cal G}$, where ${\cal G}$ is the group of Yang-Mills gauge transformations given by Equation \eqref{y-m}.  In our case the group $\G$ is a space of real smooth functions on $\Sigma$ with the group action defined as pointwise addition. The quotient $\A/\G$ can be easily described when the first de Rham cohomology of $\Sigma$ is {\em trivial}. Indeed, consider then two connections $A_1,A_2\in\A$ and their curvature two-forms $F_1,F_2$ respectively. Then $F_1=F_2$ implies that $d(A_1-A_2)=0$ and, by virtue of the triviality of the cohomology, we have  
\[
A_1=A_2+d\Omega
\]
for $\Omega\in\G$. Thus under the assumption about the first de Rham cohomology two connections  define the same element of the quotient space $\A/\G$ if and only if their curvature two-forms coincide. Assume then that $A_1,A_2$ define different points in $\A/\G$. Then there exists a point $y\in\Sigma$ such that $F_1(y)\neq F_2(y)$. Let $(y^k)$ be a local coordinate frame with the origin at $y$ such that the components $F_{1,12}(y)$ and $F_{2,12}(y)$ of the curvature two-forms are not equal. Since a holonomy $h_l(A_i)$ $(i=1,2)$ along a small (analytic) loop $l$ lying in the coordinate 'plane' given by $y^3=0$ and surrounding $y$  approximates the components the function $h_{l}$ separates the two points in $\A/\G$.        

\subsection{ACZ algebra $\sca$ and the algebra ${\cal B}$ of quantum observables \label{acz-b}}

Let us remind that the first step of the quantization strategy applied here to the 'toy theory' involves the construction of the ACZ algebra $\sca$ (Definition \ref{acz-df}) as the algebra of classical observables. At the end of the procedure we obtained the algebra $\cal B$ defined by Equation \eqref{B-alg} interpreted as the algebra of quantum observables. Now we are going to check whether the relation between the two algebras coincides with the relation between algebras of classical and quantum observables given by canonical quantization procedure. Recall that according to this procedure given algebra $\sca_0$ of classical observables is extended to a $*$-algebra $\check{\sca}_0$ of (possibly abstract) operators, next one finds a $*$-representation ${T}_0$ of $\check{\sca}_0$ on a Hilbert space. At this point one defines the algebra ${\cal B}_0$ of quantum  observables as the algebra of bounded operators on the Hilbert space. Thus if $a\in\sca_0$ defines $\check{a}\in\check{\sca}_0$ such that $\check{a}^*=\check{a}$ then ${T}_0(\check{a})$ is a self-adjoint\footnote{In fact, ${T}_0(\hat{a})$ is in general only symmetric on some dense domain but let us assume that it is also self-adjoint.} operator and consequently $\exp(it{T}_0(\check{a}))$ $(t\in\R)$ belongs to ${\cal B}_0$. Thus the canonical quantization defines the following sequence of operators:
\begin{equation}
\sca_0\ni a \mapsto \check{a} \mapsto {T}_0(\check{a})\mapsto \exp(it{T}_0(\check{a}))\in {\cal B}_0. 
\label{seq}
\end{equation}           
Now let us try to build an analogous sequence leading from the ACZ algebra $\sca$ to the algebra $\cal B$.  

Let $\check{\sca}$ be an algebra of linear operators on $\cyl$ generated by elements $\{\check{\Psi},\check{\phi}_{S,f}\}$, 
\begin{equation}
\begin{aligned}
&\cyl\,\ni\ \Psi'\mapsto\check{\Psi}\Psi'\ :=\ \Psi\Psi', \ \ \ \Psi\in\cyl,\\
&\cyl\,\ni\ \Psi'\mapsto \check{\phi}_{S,f}\Psi':=i\phi_{S,f}\Psi', \ \ \ \phi_{S,f}\in \Phi. 
\end{aligned}  
\label{repr-acz}
\end{equation}
where $\Psi$ and $\phi_{S,f}$ run through all elements of $\sca$ of this sort.  Assume now that the seemingly natural formulae
\[
\check{\Psi}^*=\check{\overline{\Psi}}, \ \ \ \check{\phi}^*_{S,f}=\check{\phi}_{S,f}
\]  
provide a well defined $*$ involution on $\check{\sca}$. Denote by $\check{\sca}_\la$ a subalgebra of $\check{\sca}$ generated by $\sca_\la$ which means, in particular, that $\check{\sca}_\la$ preserves the space $\Cyl^S_{\tL=\la}$. Let ${\cal J}_\la$ be a (right and left) ideal in $\check{\sca}_\la$ defined as 
\[
{\cal J}_\la:=\{\ \hat{a}\in\check{\sca}_\la\ | \ \hat{a}\,\Cyl^S_{\tL=\la}=0\ \}.
\]
Then it is easy to see that
\begin{equation}
\hat{\sca}_\la=\check{\sca}_\la/{\cal J}_\la,
\label{J-red}
\end{equation}
where $\hat{\sca}_\la$ is the $*$-algebra defined by \eqref{repr}. 

Now we can write down the following sequence of operators:
\begin{equation}
  \sca\ni a\mapsto\check{a}\mapsto\hat{a}_\la\mapsto T_\la(\hat{a}_\la)\mapsto\exp (it T_\la(\hat{a}_\la))\mapsto \mathbb{A}^t_\la\in {\cal B},
\label{seq-1}
\end{equation}
which should be understood as follows: an element $a$ of $\sca$ generates $\check{a}\in\check{\sca}$. Assuming that $a\in\sca_\la$ we can project (see \eqref{J-red}) $\check{a}\in\check{\sca}$ onto $\hat{a}_\la\in\hat{\sca}_\la$. If $\hat{a}_\la=\hat{a}^*_\la$ then $T_\la(\hat{a}_\la)$ is a self-adjoint operator on $\h_\la$ which defines a unitary operator $\exp (it T_\la(\hat{a}_\la))\in{\cal B}_\la$. The latter defines via the inductive limit \eqref{B-alg} the operator $\mathbb{A}^t_{\la}$ as an element of ${\cal B}$.  

Comparing the sequences \eqref{seq} and \eqref{seq-1} we see that, in the case of canonical quantization the classical observable $a$ is associated with exactly one quantum observable $\exp(itT_0(\check{a}))$ (or, more precisely, with exactly one family of quantum observables). In the case of the 'toy theory' if $a$ belongs to $\sca_{\la_0}$ then it does to any $\sca_\la$ such that $\la\geq\la_0$. Consequently, the operator $\check{a}$ undergoes many distinct projections given by distinct ideals $\{{\cal J}_\la\}$ resulting in many $\{\hat{a}_\la\}$ which, in general, are also distinct as we will see soon. It seems then that in this case the classical observable $a$ is associated with many distinct quantum observables $\mathbb{A}^t_\la$. The question now is whether the operators $\{\mathbb{A}^t_{\la}\}$ (or  $\{T_\la(\hat{a}_\la)\}$) can be combined in a way into a single operator which could be evaluated on $\D$ (or on a subset of $\D$) and be interpreted as a quantum counterpart of $a$.

The answer to that question is negative as we will show by setting $a=\phi_{S,f}$. Let us start by rewriting the sequence \eqref{seq-1}:
\begin{equation}
\sca\ni \phi_{S,f} \mapsto\check{\phi}_{S,f}\mapsto\hat{\phi}_{S,f,\la}\mapsto T_\la(\hat{\phi}_{S,f,\la})\mapsto\exp (it T_\la(\hat{\phi}_{S,f,\la}))\mapsto {\Phi}^t_{S,f,\la}\in {\cal B}.
\label{seq-phi}
\end{equation}     
Note that any flux operator $\phi_{S,f}$ belongs to all subalgebras $\{\sca_\la\}$ hence the r.h.s. of the above sequence is well defined for every $\la$. 

Given $\la=\tL$, the ideal ${\cal J}_\la$ contains inter alia all linear combinations of those flux operators which vanish on holonomies along hoops belonging to $\tL$ but whose action on holonomies along the other hoops are arbitrary. The consequence is that the operator $\hat{\phi}_{S,f,\la}$ can 'see' only hoops being elements of $\tL$ and is 'blind' with respect to the others. It is clear now that for $\la\neq\la'$ we have ${\cal J}_\la\neq{\cal J}_{\la'}$ and, in general,  
\[
\hat{\phi}_{S,f,\la}\neq\hat{\phi}_{S,f,\la'}.      
\]

Now let us focus our attention on the operator $T_\la(\hat{\phi}_{S,f,\la})$. It is  unbounded and (essentially) self-adjoint and its (dense) domain is $\pr^{*-1}_{\tL}\Cyl^S_{\tL}\subset\h_\la$ ($\la=\tL$). Consequently, the operator cannot be evaluated on the whole $\D_\la$. We define then $\D^S_\la\subset\D_\la$ as a set of {\em finite} convex combinations of density matrices of the form
\[
\ket{\Psi}\bra{\Psi}, \ \ \Psi\in\pr^{*-1}_{\tL}\Cyl^S_{\tL}.
\]   
Evidently, evaluation of $T_\la(\hat{\phi}_{S,f,\la})$ on $\D^S_\la$ provides a finite result\footnote{For $\rho_\la\in\D^S_\la$ we have
\[
|\,\langle T_\la(\hat{\phi}_{S,f,\la})\,,\rho_\la\rangle\,|=|\,\sum_{i=1}^n \scal{\Psi_i}{T_\la(\hat{\phi}_{S,f,\la})\,\Psi_i}_\la\,|<\infty,
\]              
where $\scal{\cdot}{\cdot}_\la$ is the scalar product on $\h_\la$, and $\{\Psi_i\}$ are vectors in the Hilbert space.}. Define
\[
\D^S:=\{\ \rho=\{\rho_\la\}\in\D \ | \ \rho_\la\in\D^S_\la \ \text{for all $\la\in\Lambda$}\ \}
\]
and consider $\rho\in\D^S$. Then, in general, for $\la'\neq\la$  
\[
\langle T_{\la'}(\hat{\phi}_{S,f,\la'}),\rho_{\la'}\rangle\neq\langle T_\la(\hat{\phi}_{S,f,\la}),\rho_\la\rangle.
\]   
This inequality holds even if $\la'\geq\la$---the reason for this is twofold: $(i)$ the projection from $\rho_{\la'}$ to $\rho_{\la}$ annihilates some non-trivial information contained in $\rho_{\la'}$ and $(ii)$ the operator $T_\la(\hat{\phi}_{S,f,\la'})$ can 'see' the hoops belonging to $\tL'=\la'$, while $T_\la(\hat{\phi}_{S,f,\la})$ can 'see' those belonging to $\tL=\la$ only. Thus the operators $\{T_\la(\hat{\phi}_{S,f,\la})\}$ cannot be combined into a single one which can be consistently evaluated on $\D^S$. 
 
As the last resource, we can try to combine $\{T_\la(\hat{\phi}_{S,f,\la})\}$ into a single operator by means of the limit
\begin{equation}
\lim_\la\,\langle T_\la(\hat{\phi}_{S,f,\la}),\rho_\la\rangle.
\label{lim-la}
\end{equation}
However, this limit do not exists. To convince ourselves about that consider a Schwarz function $\psi$ on $\R$ such that 
\begin{equation}
\int_{\R} \bar{\psi}\psi\,dx=1\ \ \text{and}\ \ \int_{\R} \bar{\psi}\partial_x\psi\,dx\neq0.
\label{int-psi}
\end{equation}
Given almost analytic loop $l$, choose its orientation in such a way that
\[
-2\phi_{S,f}h_l=\check{f}_l\geq 0.
\] 
Next, by means of ${\cal I}_{\{l\}}$ map the density matrix $\ket{\psi}\bra{\psi}$ to one in $\D_{\tL(\tll)}$ \footnote{The symbol $\D_{\tL(\tll)}$ was introduced in Subsection \ref{diff-inv-ISQ}.} which will be denoted by $\rho(\tL(\tll))$.  Combine all $\{\rho(\tL(\tll))\}$ into a state $\rho\in\D^S$. By direct calculation we get
\begin{equation}
\langle T_\la(\hat{\phi}_{S,f,\la}),\rho_\la\rangle=-\frac{1}{2}[\sum_{I=1}^N \check{f}_{l_I}]\int_{\R} \bar{\psi}\partial_x\psi\,dx,
\label{no-bound}
\end{equation}
where the loops $\{l_1,\ldots,l_N\}$ generate the group $\tL=\la$. It is clear now that in this case the limit \eqref{lim-la} is divergent unless $\phi_{S,f}=0$.  

The above discussion can be summarize as follows

\begin{cor}
There is no natural way to evaluate  any flux operator $\phi_{S,f}$ on $\D^S$. Instead there is a net of operators $\{T_\la(\hat{\phi}_{S,f,\la})\}$ each of them can be evaluated on $\D^S_\la$.     
\label{no-flux}
\end{cor}

Similarly, the unitary operators    
\begin{equation}
\{\ \exp(\,it \,T_\la(\hat{\phi}_{S,f,\la})\,)\ \}
\label{exp-phi}
\end{equation}
belonging respectively to $\{{\cal B}_\la\}$ do not define any single operator in $\cal B$. To see this consider $\la'\geq\la$. Then there using the embedding $\pi^*_{\la\la'}$ introduced in Theorem \ref{B-pi-thr} we get
\begin{equation}
\pi^*_{\la\la'}\exp(\,it \,T_\la(\hat{\phi}_{S,f,\la})\,)=\id\ot u^{-1}_{\la'\la} \exp(\,it \,T_\la(\hat{\phi}_{S,f,\la})\,),
\label{pi-star}
\end{equation}
where $\id$ is the identity operator on the Hilbert space $\tl{\h}_{\la'\la}$ given by the decomposition \eqref{hhh}, and $u_{\la'\la}$ is defined by \eqref{ull}. In general, the r.h.s. of \eqref{pi-star} do not coincide with $\exp(\,it \,T_{\la'}(\hat{\phi}_{S,f,\la'})\,)$. In other words, the net \eqref{exp-phi} of exponentiated operators does not define any single element of $\cal B$ but each of them defines via the inductive limit a distinct operator
\begin{equation}
\Phi^t_{S,f,\la}=\underrightarrow{\lim} \ \pi^*_{\la'\la}\,\exp(\,it \,T_\la(\hat{\phi}_{S,f,\la})\,)\in{\cal B}.
\label{Phi-net}
\end{equation}
appearing at the end of the sequence \eqref{seq-phi}.

One can still ask whether the net $\{\Phi^t_{S,f,\la}\}$  is convergent in a topology on $\cal B$. Here it is natural to consider two topologies: the norm topology i.e. one defined by the $C^*$-algebra norm on $\cal B$) and the weak topology. The convergence of the net $\{\Phi^t_{S,f,\la}\}$ in the weak topology would mean that the net
\begin{equation}
\langle \Phi^t_{S,f,\la},\rho\rangle=\langle\exp(\,it \,T_\la(\hat{\phi}_{S,f,\la})\,),\rho_\la\rangle.
\label{Phi-rho-net}
\end{equation} 
is convergent for every $\rho\in\D$. The answer to the question we are able to give now is not complete and can be formulated as the following lemma which is proven in Appendix \ref{net-app}.
\begin{lm}
Suppose $\phi_{S,f}\neq 0$. Then 
\begin{enumerate}
\item the corresponding net $\{\Phi^t_{S,f,\la}\}$ is divergent in the norm topology.
\item For every $\rho\in\D$  the net \eqref{Phi-rho-net} is bounded
\[
|\langle \Phi^t_{S,f,\la},\rho\rangle|\leq 1.
\]
Moreover, the equality holds if and only if $\Phi^t_{S,f,\la}$ is the unit of $\cal B$.  
\item If $\rho=\{\rho_\la\}$ is such that for every $\la\in\Lambda$
\[
\rho_\la=\rho(\tL(l_1))\ot\ldots\ot \rho(\tL(l_N)),
\]  
where $\{l_1,\ldots,l_N\}$ are independent loops generating $\tL=\la$, $\tL(l_I)$ is a tame group generated by the loop $l_I$ and $\rho(\tL(L_I))\in D_{\tL(l_I)}$, then
\begin{equation}
\lim_\la\,\langle \Phi^t_{S,f,\la},\rho\rangle =0.
\label{lim-zero}
\end{equation}
\end{enumerate}
\label{net-lm}
\end{lm}
Although we have to leave open the issue of the weak convergence of the net $\{\Phi^t_{S,f,\la}\}$ the result \eqref{lim-zero} suggests that the net may be convergent to zero.

The discussion presented above can be summarized by saying that the relation between the ACZ algebra algebra and ${\cal B}$ seems to be essentially different from the relation between the algebras of classical and quantum observables obtained by applying canonical quantization: the sequences \eqref{seq} and \eqref{seq-1} appear to be not reconcilable.

Let us finally note that above we were considering some operators evaluated on the spaces $\D$ and $\D^S$ of states which are only Yang-Mills gauge invariant. Let us check what would happen if we restricted ourselves to diffeomorphism invariant states in $\D_{\rm ph}$ and $\D^S\cap\D_{\rm ph}$. While Lemma \ref{net-lm} is insensitive to this restriction (see its proof in Appendix \ref{net-app}) Corollary \ref{no-flux} has to be reformulated. The reason is that now the states $\{\rho(\tL(\tll))\}$ used to derive Equation \eqref{no-bound} cannot be chosen arbitrarily since there can exist a diffeomorphism $\chi$ such that $\chi(\tll)=\tll^{-1}$. Then the diffeomorphism would induce an action on every $\D_{\tL(\tll)}$ and every $\{\rho(\tL(\tll))\}$ would have to be invariant with respect to the action. To ensure the invariance it is enough to require that the function $\psi$ defining $\{\rho(\tL(\tll))\}$  satisfies $\psi(-x)=e^{it}\psi(x)$ for some real $t$ (the map $x\mapsto -x$ on $\R$ corresponds to the change of the orientation $\tll$ caused by $\chi$). Such a $\psi$, however, does not satisfy the second requirement of \eqref{int-psi}. What we can do in this situation is to replace that requirement by
\[
\int_{\R} \bar{\psi}\partial^2_x\psi\,dx\neq0
\]       
and show that the net
\begin{equation}
\lim_\la\,\langle T_\la(\hat{\phi}^2_{S,f,\la}),\rho_\la\rangle.
\label{lim-la-2}
\end{equation}
is also divergent. Hence we have

\begin{cor}
There is no natural way to evaluate the $\check{\phi}^2_{S,f}\in \check{\sca}$ on $\D^S\cap\D_{\rm ph}$. Instead there is a net of operators $\{T_\la(\hat{\phi}^2_{S,f,\la})\}$ each of them can be evaluated on $\D^S_\la$.     
\end{cor}
 
\subsection{Comparison with the compact case \label{comp}}

At this point we are able to perceive some important differences between the quantum 'toy theory' and a quantum model in the compact case like e.g. LQG. Let us begin by analyzing the space of quantum states. 

In the compact case the space of (pure) quantum states is the Hilbert space $\h=L^2(\Abar,d\mu_{\rm AL})$, where $\Abar$ is the Ashtekar-Isham configuration space of generalized connections \cite{ai,proj} and $d\mu_{\rm AL}$ is the Ashtekar-Lewandowski measure \cite{al-hoop} on $\Abar$. The Hilbert space $\h$ can be obtained equivalently by means of the inductive limit \eqref{h}. On the other hand the space $\D$ of quantum states of the 'toy theory' is defined as the projective limit \eqref{D-df} and is not a Hilbert space.

 Let us now emphasize another essential difference between the spaces $\D$  and $\h$. Given a tame hoop group $\tL=\la$, in the former case there exists a canonical projection $\pi_{\la}:\D\rightarrow\D_\la$, while in the latter one, there exists a canonical embedding $p_\g:\h_\g\rightarrow\h$ where $\g$ is a graph. Since any projection is equivalent to a loss of some information any state $\rho_\la:=\pi_\la\rho$ belonging to $\D_\la$ should be considered only as an {\em approximation} of the state $\rho\in\D$ no matter how large and complicated the group $\tL$ is. On the other hand, every state $v_\g\in\h_\g$ is naturally identified with a state $v:=p_\g v_\g\in\h$, i.e. it is just a state, no matter how trivial the graph $\g$ is. Thus to use the projective techniques to build the space of quantum states means to construct the space by a sequence (or, more precisely, by a net) of approximations.  

Note moreover that every vector in $\h$ is a sum of at most countable cylindrical functions (like e.g. spin-networks---see \cite{rov,rev})) and every cylindrical function depends on holonomies along {\em finite} number of edges. Consequently, every vector can be seen as a function depending on at most {\em countable} number of degrees of freedom represented by the holonomies. We can state this equivalently: every state depends on all the degrees of freedom but for most of them the dependence is trivial. 

This is, however, not true in the case of the quantum 'toy theory': given loop $l$, every state $\rho\in\D$ can be projected to $\rho_\la$, where $\la=\tL(l)$ is the tame group generated by $l$. Since $\rho_\la$ is a density matrix on the Hilbert space $\h_\la$ defined as a space of $L^2$ functions on a {\em non-compact} space $\A_{\tL(l)}\cong\R$ it cannot be represented by $\ket{\Psi}\bra{\Psi}$ where $\Psi$ is a constant function on $\A_{\tL(l)}$. Thus $\rho_\la$ does depend non-trivially on the degree of freedom defined by the holonomy along the loop $l$. Consequently, every state $\rho$ depends non-trivially on {\em all} the (configuration) degrees of freedom which constitute a set of {\em uncountable} cardinality. This conclusion should not be surprising: in quantum mechanics we define the space of quantum states as a set of square-integrable functions on a non-compact configuration space and then each function depends non-trivially on all degrees of freedom. Here we just obtained an analogous result.

Now let us focus our attention on Equation \eqref{e-psi-braket} which (together with Equation \eqref{hat-phi}) defines a flux operator. Given space $\Cyl_{\tL}$, the flux operator $\phi_{S,f}$ is defined as a finite sum of derivation operators such that each of them corresponds to a degree of freedom defined by the holonomy along a loop. Denote by $\phi_{S,f,l}$ such a derivation operator associated with the loop $l$. Then $\phi_{S,f}$ as an operator on the whole $\Cyl$ can be expressed as     
\begin{equation}
\phi_{S,f}=\sum_{\tL(l)} \phi_{S,f,l}\,,
\label{flux-l}
\end{equation}
where the sum runs through all tame groups generated by single loops. Thus the flux operator is given by an {\em uncountable} sum of operators corresponding to single degrees of freedom. In the compact case flux operators can also be expressed similarly as an uncountable sum (see e.g. \cite{rev}). 

We can see now why it is possible to define the action of flux operators on $\h$ (i.e. in the compact case), and why it is impossible to evaluate flux operators  on $\D$ (i.e. in the non-compact case)---in the former case a state depends non-trivially only on at most countable number of degrees of freedom, so each time the flux operator acts on a state the uncountable sum in \eqref{flux-l} reduces automaticly to a countable one. In the latter case this is not true since the states in $\D$ do depend non-trivially on all the degrees of freedom. Thus in this case we are forced to reduce the sum by restricting ourselves to a finite number of operators $\{\phi_{S,f,l}\}$ obtaining as the result the net of the operators \eqref{Phi-net}.

The above discussion is summarized in Table \ref{table}.
\begin{table}
\begin{center}
\begin{tabular}{cccc}
& &{\bf state} &{\bf flux operator}\bigskip\\
{\bf compact case} & &countable & uncountable\medskip\\
{\bf non-compact case} & &uncountable & finite \\
\end{tabular}
\caption{Cardinality of the set of degrees of freedom on which a state and a flux operator depend non-trivially. \label{table}}
\end{center}
\end{table}

\subsection{Discretization of  geometry}

There is a common belief that quantum gravity may provide a discreet picture of the space-time instead of the continuous one assumed by general relativity. LQG partially supports this belief: quantum operators corresponding to basic geometric quantities like area and volume have discrete spectra \cite{rov,area,vol}, but still these operators are associated with regular subsets of the space-like manifold $\Sigma$ like e.g. bounded two-dimensional (analytic) submanifolds in the case of the area operator.  

In the quantum 'toy theory' the situation is quite different. We do not have, of course, any operators corresponding to classical area or volume since the classical 'toy theory' does not provide us with any metric. Therefore the only thing we can do is  to analyze the quantum counterparts \eqref{Phi-net} of classical fluxes of the momentum field across surfaces in $\Sigma$. However, Lemma \ref{net-lm} provides an evidence that $\{\Phi^t_{S,f,\la}\}$ do not define any non-trivial operator being a counterpart of  the original flux operator $\phi_{S,f}\in\sca$. If this is really the case we are left with operators which do not corresponds to the entire surface $S$ underlying $\phi_{S,f}$ since, given $\Phi^t_{S,f,\la}$,  the surface $S$ intersected originally by uncountably many loops is reduced to a finite number of points and to a finite number of loops passing through them. Thus in the 'toy theory' what seems to undergo the discretization is the surface as the set of points, while in the LQG case it is rather a feature of the surface (i.e. the area) which does not depend merely on that which points constitute the surface. We can say then that in the case of the 'toy theory' the discretization appears to proceed at the more fundamental level than in the LQG case.        

Given surface $S$ and smearing function $f$ on it, one can ask which operator \eqref{Phi-net} should be considered as one representing the classical flux across $S$. The only answer we are able to give now refers to the interpretation of a state $\rho_\la\in\D_\la$ as an {\em approximation} of $\rho=\{\rho_\la\}\in\D$. Recall that $\rho$ depends essentially on the uncountable number of configuration degrees of freedom, so the whole information encoded in it is not available for us. We are then forced to use only a finite number of measuring instruments $\{h_{l_1},\ldots,h_{l_N}\}$ which results in doing an ordinary quantum mechanics on $\A_{\tL}\cong\R^N$ where $\tL$ is generated by $\{l_1,\ldots,l_N\}$. Now, the choice of the instruments $\{h_{l_1},\ldots,h_{l_N}\}$ leads naturally to the operator $\phi_{S,f,\la}$ with $\la=\tL$, and consequently to the corresponding discretization of the surface $S$. Thus the discretization in the quantum 'toy theory' seems to be {\em observer-dependent}---it depends on how (i.e. by means of which instruments) the observer perceives the physical reality.    

Taking into account the above conclusions we cannot exclude that if someone will manage to apply the quantization method presented in this paper to GR written in terms of the $SL(2,\C)$ connections then it will be not possible to define a single area operator corresponding to a surface as a submanifold of $\Sigma$---it may happen that non-compactness of $SL(2,\C)$ will force an analogous reduction of the surface to a finite set of points. This indicates that the change of the topology of the gauge group may result in a non-trivial change of the quantum theory which may force us to rethink some important geometric notions like  the notion of a surface. In particular, it may be necessary to elaborate a new approach to the derivation of black hole entropy.

\subsection{Diffeomorphism invariant states on $\cal B$}

In the Subsection \ref{diff-inv-ISQ} we constructed explicitly quite plentiful set of diffeomorphism invariant states on the $C^*$-algebra $\cal B$. This is in sharp contrast with the compact case where there is precisely one diffeomorphism invariant state \cite{lost} on the algebra of quantum observables which is the holonomy-flux $*$-algebra $\hat{\sca}_{\rm hf}$ \cite{hanno}. Let us now point the causes which are responsible for the difference. 

Note first that $\cal B$ contains {\em exponentiated} flux-like operators $\{\Phi^t_{S,f,\la}\}$ while $\hat{\sca}_{\rm hf}$ does just flux operators. To make the comparison more transparent let us first 'de-ex\-po\-nen\-tia\-te' the operators $\{\Phi^t_{S,f,\la}\}$. Consider then an algebra of linear operators on Schwarz functions $\mathbb{S}_N$ generated by all multipliers $\mathbb{M}_N$ and derivative operators. Note that there is a natural $*$-involution on it. Given $\la=\tL$, we can use the map ${\cal I}_{\cal L}$ \eqref{bij} to pull back the algebra onto a $*$-algebra of operators acting on ${\cal I}_{\cal L}^*(\mathbb{S}_N)\subset\h_\la$. Denote this new algebra by ${\cal B}'_\la$. It is clear that $T_\la(\hat{\sca}_\la)$ restricted to ${\cal I}_{\cal L}^*(\mathbb{S}_N)$ is a subalgebra of ${\cal B}'_\la$. Given $\la'\geq\la$, taking into account Equation \eqref{pi-star} we define a map 
\[
{\cal B}'_\la \ni \hat{a}_\la\mapsto \pi^*_{\la\la'}(\hat{a}_\la):=\id\ot u^{-1}_{\la'\la} \hat{a}_\la \in {\cal B}'_{\la'},
\]      
where $\id$ is the identity operator on the Hilbert space $\tl{\h}_{\la'\la}$ given by the decomposition \eqref{hhh} and $u_{\la'\la}$ is defined by \eqref{ull}. Thus we obtained an inductive family $\{{\cal B}'_\la, \pi^*_{\la\la'}\}$ whose inductive limit ${\cal B}'$ can be thought as generated by cylindrical functions $\cyl$ and all flux-like operators $\{\hat{\phi}_{S,f,\la}\}$. Thus the algebra ${\cal B}'$ resembles $\hat{\sca}_{\rm hf}$ much more closely that $\cal B$.   

It is clear that by applying the construction of diffeomorphism invariant states presented in Subsection \ref{diff-inv-ISQ} we can obtain a large set of diffeomorphism invariant states on ${\cal B}'$. This means that  the reasoning of \cite{lost} which leads to the uniqueness of diffeomorphism invariant state on $\hat{\sca}_{\rm hf}$ does not work in the case of ${\cal B}'$. There are at least three  reasons for this:
\begin{enumerate}
\item in \cite{lost} diffeomorphism invariance of a state $\omega$ on $\hat{\sca}_{\rm hf}$ is defined as the invariance with respect to all {\em semi-analytic} diffeomorphisms on $\Sigma$ which form a larger group that analytic diffeomorphisms considered in this paper;
\item the crucial step of the proof of the uniqueness theorem of \cite{lost} shows that for every flux operator ${X}_{S,f}$ 
\begin{equation}
\omega(\hat{X}^*_{S,f}\hat{X}_{S,f})=0.
\label{omega-X-0}
\end{equation}
This result is obtained by observing that the expression $\omega(\hat{X}^*_{S,f}\hat{X}_{S,f'})$ defines a {\em diffeomorphism invariant} scalar product between functions on $S$ and by showing that the invariance forces the product to be zero. This reasoning cannot be applied to ${\cal B}'$ since the flux-like operators $\{\hat{\phi}_{S,f,\la}\}$ are not associated with surfaces but with finite subsets of $\Sigma$.   
\item in \cite{lost} the cylindrical functions are defined over graphs embedded in $\Sigma$. In this case edges of a graph can be considered as {\em compositions} of edges of another one i.e. we can transform a (smaller) graph into another (bigger) one by splitting its edges. In the case of ${\cal B}'$ the cylindrical functions are defined over loops (as a special kind of graphs) which are {\em almost analytic}. As stated by Lemma \ref{almost+-} no almost analytic loop can be a composition of other such loops. This property is also a reason of plentitude of diffeomorphism invariant states on $\cal B'$.

To see this assume that the loops are just piecewise analytic and consider edges $\{e_0,e_1,e_2\}$ sharing the same beginning point and the same final one. Suppose also that there exists a diffeomorphism $\chi$ such that $\chi(e_i)=e_{i+1|{\rm mod}\, 3}$ and define loops
\[
l_{0}:=e^{-1}_1\circ e_2, \ \ l_{1}:=e^{-1}_2\circ e_0, \ \ l_{2}:=e^{-1}_0\circ e_1. 
\]
Let $\rho$ be a diffeomorphism invariant state constructed by the method presented in Subsection \ref{diff-inv-ISQ}. Consequently, each $\rho(\tL(l_i))$ ($i=1,2,3$) is defined by a density matrix $\rho([l_i])$ on $L^2(\R,dx)$. Since $\chi(l_i)=l_{i+1|{\rm mod}\, 3}$ the matrix $\rho([l_i])$ does not depend on $i$. Because $l^{-1}_0=l_1\circ l_2$ for every bounded function $\psi$ on $\R$ understood as a bounded operator on $L^2(\R,dx)$ there holds the following condition  
\begin{equation}
\langle \,\rho([l_i])\,,\,\kappa \psi\,\rangle=\langle\,\rho([l_i])\ot\rho([l_i])\,,\,\Delta\psi\,\rangle, 
\label{restr}
\end{equation}
where 
\[
(\kappa\psi)(x)=\psi(-x) 
\]
and $\Delta\psi$ is a bounded function on $\R^2$ given by
\[
(\Delta\psi)(x_1,x_2)=\psi(x_1+x_2).
\]  

We thus see that if we allowed the loops to be decompositions of other ones the number of diffeomorphism invariant states constructed in Subsection \ref{diff-inv-ISQ} would be considerably reduced by the condition \eqref{restr}. However, this condition is not valid when the loops are almost analytic. Note also, that the set of almost analytic loops is not preserved by semi-analytic diffeomorphisms used in \cite{lost} and this is the main reason why in this paper we use only analytic ones.
\end{enumerate}  
 
We conclude that $(i)$ application of analytic diffeomorphisms, $(ii)$ the discretization of the surfaces in the case of the flux-like operators and $(iii)$ application of almost analytic loops make the diffeomorphism invariance much less restrictive than in the case considered in \cite{lost} and therefore the set of diffeomorphism invariant states on ${\cal B}'$, and consequently, on $\cal B$ is very large.

\paragraph{Acknowledgments} I am very grateful to Prof. Jerzy Kijowski for the idea and discussions. I also would like the thank Prof. Abhay Ashtekar, Wojciech Kami\'nski, Jerzy Lewandowski, Jorge Pullin and Thomas Thiemann for discussions and Jacek Jezierski, Kazimierz Napi\'orkowski and Wies{\l}aw Pusz for their help with the proof of Lemma \ref{net-lm}. The author was partially supported by the Polish Ministerstwo Nauki i Szkolnictwa Wy\.zszego grant 1 P03B 075 29.

\appendix

\section{Physical degrees of freedom of the 'toy theory' \label{deg-free}}

Here we are going to construct an injective map from the set $]0,1[\times\R$ into the class \eqref{con-sol} of solutions of the constraints such that
\begin{enumerate}
\item given two distinct pairs $(\alpha,\beta),(\alpha',\beta')\in]0,1[\times\R$ there is no gauge transformation which maps the fields corresponding to $(\alpha,\beta)$ to ones corresponding to $(\alpha',\beta')$;   
\item given $\alpha\in]0,1[$, the fields depend continuously on $\beta\in\R$.  
\end{enumerate}

Assume that $\Sigma=\R^3$. Consider the cylindrical coordinates $(\rho,\phi,z)$ on it and a smooth function on $\Sigma$ of the form 
\[
b(\rho,z)=a(\rho)\sin(\rho) \exp(-z^2),
\]  
where $a(\rho)$ is a non-negative smooth function on $[0,\infty]$ such that it is zero on $[0,2\pi]$ and is strictly positive on $[2\pi,\infty]$. We assume moreover that for large $\rho$'s the function behaves like a Schwarz function e.g. like $\exp(-\rho^2)$. Note that the function $b(\rho,z)$ distinguishes infinite 'rings' in $\Sigma$ such that on each ring the function is either (strictly) positive or negative. 

Consider now a smooth non-negative function $\xi_{n,m}$ ($n=1,2,\ldots$, $m=0,1,\ldots,9$) on $[0,\infty]\times\R$ such that $(i)$ its support is a ball of the center at $(2\pi+\frac{\pi}{2}n,m)$ and the radius $\epsilon\ll 1$ and $(ii)$ the maximum value of $\xi_{n,m}$ is $1$  and is reached at precisely one point.

Given $\alpha\in]0,1[$, let $\alpha_n$ be the $n$-th digit of the decimal expansion of $\alpha$. Define
\[
\zeta_\alpha:=1-\sum_{\substack{n=1\\\alpha_n\neq0}}^\infty\sum_{m=1}^{\alpha_n}\xi_{n,m}.
\]     
Evidently, $\zeta_\alpha$ is smooth, non-negative and not greater than $1$.

Note now that the function
\[
f_\alpha(\rho,\phi,z):=b(\rho,z)\zeta_\alpha(\rho,z)
\] 
encodes the full information about the number $\alpha$ in its zero set: in the $n$-th 'ring' there are precisely $\alpha_n$ pairwise distinct circles on which the value of the function is equal to $0$.   

Define on $\Sigma$ a connection one-form 
\[
A_{\alpha,i}\,dx^i:=[\int_{0}^z f_\alpha(\rho,\phi,z')\, dz']\, d\rho
\]  
and a vector density
\begin{equation}
\tl{E}^i_{\alpha,\beta}:=\beta \,\tl{\epsilon}^{ijk}F_{\alpha,jk},
\label{E-F}
\end{equation}
where $\beta$ is a real number, and 
\[
F_{\alpha,ij}\,dx^i\wedge dx^j=f_\alpha(\rho,\phi,z)\,dz\wedge d\rho
\] 
is the curvature form of $A_{\alpha}$. It is easy to check that $(\tl{E}^i_{\alpha,\beta},A_{\alpha,j})$ is a solution of all the constraints (see \eqref{sol-E} and \eqref{sol-g}). 

Consider now two pairs $(\tl{E}^i_{\alpha,\beta},A_{\alpha,j})$ and $(\tl{E}^i_{\alpha',\beta'},A_{\alpha',j})$ and assume that there exist a diffeomorphism $\chi$ and a Yang-Mills gauge transformation such that they map the former pair to the latter one. This means in particular that 
\[
(\chi^{-1})^*F_{\alpha}=F_{\alpha'}. 
\]
This equation implies that the zero set of $F_\alpha$ is mapped bijectively onto the zero set of $F_{\alpha'}$. However, it is easy to realize that the zero set of $F_\alpha$ transformed by any diffeomorphism still encodes the full information about $\alpha$. Hence $\alpha=\alpha'$ and $F_\alpha=F_{\alpha'}$. It follows from \eqref{E-F} that
\[
\beta=\beta'.
\]

This means that 
\[
(\alpha,\beta)\mapsto(\tl{E}^i_{\alpha,\beta},A_{\alpha,j})
\]
is an injective map from $]0,1[\times\R$ into the set of the gauge orbits contained in the physical phase space. Note moreover that, given $\alpha$, the fields $(\tl{E}^i_{\alpha,\beta},A_{\alpha,j})$ depend continuously on $\beta$. Therefore we can say that $\beta$ parameterizes the degree of freedom corresponding to a fixed $\alpha$. Since the interval $]0,1[$ is an uncountable set the 'toy theory' possesses an uncountable number of physical degrees of freedom.

\section{Dependence of $\pi_{\la\la'}$ on the choice of $\A_{\la'\la}$ \label{pi-A} \label{dep-app}}

Here we are going to show that the projection $\pi_{\la\la'}:\D_{\la'}\rightarrow\D_\la$ depends essentially on the choice of $\A_{\la'\la}$ in the decomposition \eqref{a-dec}. Recall that \eqref{a-dec} generates the decomposition of the Hilbert space $\h_{\la'}$ into the tensor product $\tl{\h}_{\la'\la}\ot\h_{\la'\la}$ and $\pi_{\la\la'}$ is defined by evaluating the partial trace with respect to $\tl{\h}_{\la'\la}$ and by identifying the resulting density matrix on $\h_{\la'\la}$ with one on $\h_\la$ (to identify the matrices we use a natural unitary map between the two Hilbert spaces). In fact, this is the partial trace which is sensitive to the choice of $\A_{\la'\la}$. We will show this by presenting a very simple example. 

Let $\R^2$ $(\R)$ play the role of $\A_{\tL'}$ $(\A_{\tL})$ and let the projection
\[
\R^2\ni(x,y)\mapsto\pr(x,y):=x\in\R
\]   
play the role of $\pr_{\tL\tL'}$. Clearly, $\ker\pr$ is the $y$-axis $Y$ of $\R^2$. Given Hilbert spaces
\[
L^2(\R^2,dxdy) \ \ \text{and} \ \ L^2(\R,dx),
\]
we want to find a projection $\pi$ which corresponds to the projection $\pr$ and  maps density matrices on $L^2(\R^2,dxdy)$ to ones on $L^2(\R,dx)$.

Let us choose a subspace $\R_a$  of $\R^2$ spanned by a vector $(1,a)$ as the space $A_{\la'\la}$. Then the map
\[
\R\ni x\mapsto \theta_a(x):=(x,ax)\in\R_a\subset\R^2
\]
satisfies $\pr\circ\theta_a=\id$ (and hence is a counterpart of the map $\theta_{\la'\la}$ introduced in Subsection \ref{pre}). Note that $Y$ as $\ker\pr$ and $\R_a$  give us a decomposition of $\R^2$ which is a counterpart of \eqref{a-dec}. 

Now we are going to decompose the Hilbert space $L^2(\R^2,dydx)$. Let $(x',y')$ be coordinates on $\R^2$ given by
\[
\begin{cases}
x'=x &\\
y'=y-ax&
\end{cases}.
\]   
Evidently, the condition $x'=0$ describes the axis $Y$ while $y'=0$ does $\R_a$. Consequently, $x'$ restricted to $\R_a$ is a coordinate on it while $y'$ restricted to $Y$ is a coordinate on the axis. It is straightforward to check that 
\begin{equation}
x'(\theta_a(x))=x,
\label{x'-th-x}
\end{equation}
where $x'$ is meant to be the coordinate on $\R_a$. Moreover,
\[
dydx=dy'dx' \ \ \text{and} \ \ dx'=\theta_{a*}dx
\]    
where $dx'$ $(dx)$ is understood here as a measure on $\R_a$ $(\R)$. Hence we get the desired decomposition of the Hilbert space $L^2(\R^2,dydx)$, 
\[
L^2(\R^2,dydx)=L^2(Y,dy')\ot L^2(\R_a,dx').
\] 

Consider a density matrix $\rho$ on $L^2(\R^2,dxdy)$. As it was said above the projection $\pi$ should be defined by evaluating the partial trace $\tr_Y$ of $\rho$ with respect to $L^2(Y,dy')$ and then by mapping the resulting density matrix $\tr_Y\rho$ on $L^2(\R_a,dx')$ onto a density matrix $\pi\rho$ on $L^2(\R,dx)$. To evaluate the partial trace $\tr_Y$ we represent $\rho$ by its integral kernel 
\[
\rho(\tl{x},\tl{y},x,y)=\rho(\tl{x}',\tl{y}'+a\tl{x}',x',y'+ax').
\]
Then the kernel of $\tr_Y\rho$ is given by 
\begin{multline}
(\tr_Y\rho)(\tl{x}',x')=\int_\R \rho(\tl{x}',y'+a\tl{x}',x',y'+ax')\,dy'=\\=\int_\R \rho(\tl{x}',y',x',y'+a(x'-\tl{x}'))\,dy'.
\label{tr-1}
\end{multline}
We see then that $\tr_Y\rho$ depends essentially on the choice of $\R_a$. This dependence does not change after passing from $\tr_Y\rho$ to $\pi\rho$. It is clear that  $\theta_a^*$ defines the natural unitary map from $L^2(\R_a,dx')$ onto $L^2(\R,dx)$, hence the passage from $\tr_Y\rho$ to $\pi\rho$ is given by this map. Applying \eqref{x'-th-x} we obtain the formula describing the kernel of $\pi\rho$,
\[
(\pi\rho)(\tl{x},x)=\int_\R \rho(\tl{x},y',x,y'+a(x-\tl{x}))\,dy',
\] 
which still depends on the choice of $\R_a$.

\section{Proof of Lemma \ref{N'>N} \label{app-proof}}

Lemma \ref{N'>N} follows immediately from:
\begin{lm}
Suppose that a tame group $\tL$  generated by the set $(l_1,\ldots,l_N)$ of independent loops is a subgroup of a tame group $\tL'$ generated by the set $(l'_1,\ldots,l'_{N'})$ of independent loops. Then $(i)$ $\tL'=\tL$  if and only if $N'=N$ and $(ii)$ $\tL$ is a proper subgroup of $\tL'$ if and only if $N'>N$.        
\label{LL'}
\end{lm}

\noindent{\it Proof.} Let $\tL'$ and $\tL$ be the tame groups of hoops occurring in Lemma \ref{LL'} generated respectively by sets ${\cal L}'=\{l'_1,\ldots,l_{N'}\}$ and ${\cal L}=\{l_1,\ldots,l_{N}\}$ of independent loops.  Applying a construction described in \cite{al-hoop} we can get a set ${\cal K}=\{{k}_1,\ldots,{k}_n\}$ of independent loops  such that:
\begin{enumerate}
\item  every loop belonging to ${\cal L}'\cup{\cal L}$ is a composition of loops belonging to ${\cal K}$,
\item given a loop $l\in{\cal L}'\cup{\cal L}$ there exists a loop $k_i\in{\cal K}$ such that  
\begin{enumerate}
\item the loop $l$  can be decomposed as $l= k'_1\circ{k_i}^{\pm 1}\circ k'_2$, where $k'_1, k'_2$ are loops built from the ones in ${\cal K}$ {\em except} ${k}_i$. Without loss of generality we will assume that the  orientation of the loop  $l$ is such that $l= k'_1\circ k_i\circ k'_2$.     
\item if the loop $l\in {\cal L}'\,({\cal L})$ then it is the only loop in ${\cal L}'\,(\cal L)$ in whose decomposition the loop ${k_i}$ appears.
\end{enumerate}
\end{enumerate}

Now, decompose each loop $l\in{\cal L}'\cup{\cal L}$ in terms of the loops $\{{k}_1,\ldots,{k}_n\}$. Thus we obtain the following decomposition of hoops:  
\begin{equation}
\begin{aligned}
&\tilde{l}'_{I'}=\prod_{i=1}^{n}(\tilde{{k}}_i)^{M'_{I' i}}:=(\tilde{{k}}_1)^{M'_{I' 1}}\circ\ldots\circ(\tilde{{k}}_n)^{M'_{I' n}},\\
&\tilde{l}_{I}=\prod_{i=1}^{n}(\tilde{{k}}_i)^{M_{I i}}:=(\tilde{{k}}_1)^{M_{I 1}}\circ\ldots\circ(\tilde{{k}}_n)^{M_{I n}}.
\end{aligned}
\label{matrix}
\end{equation}
where every component of the matrices $M'=(M_{I' i})$ and $M=(M_{I i})$ belongs to $\Z$. Clearly, the matrix $M'$ $(M)$ has  $n$ columns and $N'$ $(N)$ rows. The properties of the decomposition of the loops in ${\cal L}'$ $({\cal L})$ in terms of the loops in $\cal K$  imply that the columns of $M'$ $(M)$ can be ordered in such a way that the first $N'$ $(N)$ columns of the matrix form the unit $(N'\times N')$-matrix ($(N\times N)$-matrix) \cite{oko}. This means that the rank of $M'$ $(M)$ is maximal and equal to $N'$ $(N)$. 

Since $\tL$ is a subgroup of $\tL'$ there exists a matrix $Q=(Q_{II'})$ of $N$ rows and $N'$ columns such that every component of it belongs to $\Z$ and      
\begin{equation}
\tll_I=\prod_{I'=1}^{N'} (\tll'_{I'})^{Q_{II'}}
\label{Q}
\end{equation}
Taking into account Equations \eqref{matrix} the above equation can be rewritten as
\begin{equation}
\prod_{i=1}^{n}(\tilde{{k}}_i)^{M_{I i}}=\prod_{I'=1}^{N'}\,( \prod_{i=1}^n\,(\tl{k}_i)^{M'_{I'i}})^{Q_{II'}}.
\label{kk}
\end{equation}
Because the loops in $\cal K$ are independent it is possible to find $\R$-connections $\{A_1,\ldots,A_n\}$ on $\Sigma$ such that the holonomy
\[
h_{k_i}(A_j)=\delta_{ij}.
\]     
Hence the computation of the holonomies of the connections along the loops occurring in \eqref{kk} give us by virtue of \eqref{h-lin}
\[
M_{Ii}=\sum_{I'=1}^{N'}\,Q_{II'}M'_{I'i}.
\]
For the rank of $Q$ cannot be greater than $N$ we have immediately
\begin{equation}
N'\geq N.
\label{N'-geq-N}
\end{equation}
Assume now, that $\tL'=\tL$. Then the transformation \eqref{Q} is invertible which means that the matrix $Q$ is invertible hence 
\[
N'=N.
\]   
Suppose now that the above equality holds. Then properly chosen columns of the matrix $M$ form a unit $(N'\times N')$-matrix. This means that the corresponding columns of the matrix $M'$ constitute the matrix $Q^{-1}$ and that the components of $Q^{-1}$ belong to $\Z$. Hence
\[
\prod_{I=1}^{N'} (\tll_I)^{Q^{-1}_{I'I}}=\tll'_{I'}
\]
which means that $\tL'=\tL$. Thus we have proved the statement $(i)$ of the lemma. This statement together with \eqref{N'-geq-N} imply the statement $(ii)$. $\blacksquare$

\section{Proof of Lemma \ref{net-lm} \label{net-app}}

Let us fix a surface $S$ and a non-zero function $f$ on it. Then $\phi_{S,f}\neq 0$. In order to prove Statement 1 of the lemma note that for every $\la'\geq\la$ 
\begin{equation}
||\Phi^t_{S,f,\la}||=||\id\ot u^{-1}_{\la'\la} \exp(\,it \,T_\la(\hat{\phi}_{S,f,\la})\,)||,
\label{norms}
\end{equation}
where the equality comes from Equation \eqref{pi-star} and the norm on the r.h.s. of the formula is the norm on ${\cal B}_{\la'}$. Recall that $\h_{\la'}\cong L^2(\R^{N'},dx)$, where the isomorphism between the Hilbert spaces is given by the map \eqref{bij}. Using the isomorphism we can convince ourselves that the operator on the r.h.s. of \eqref{norms} corresponds to a translation operators ${\cal T}_{x}$ $(x\in\R^{N'})$, 
\[
({\cal T}_{x}\psi)(x'):=\psi(x'+x), 
\]
on $L^2(\R^{N'},dx)$. Consider now two operators ${\cal T}_{x}$ and ${\cal T}_{x'}$ such that $x\neq x'$ and a function $\psi\in L^2(\R^{N'},dx)$ of compact support such that the support is contained in a ball of the radius not greater than $||x-x'||/2$. Then we have
\[
\scal{({\cal T}_{x}-{\cal T}_{x'})\psi}{({\cal T}_{x}-{\cal T}_{x'})\psi}=2\scal{\psi}{\psi}
\]        
which means that 
\begin{equation}
||{\cal T}_{x}-{\cal T}_{x'}||\geq 2. 
\label{TT}
\end{equation}

Let $\tL=\la$ be an arbitrary tame hoop group. There exists a loop $l$ such that $\tll\not\in\tL$ and $\phi_{S,f}h_l\neq 0$. Let $\tL'=\la'\geq\la$ be a tame group generated by hoops in $\tL$ and $\tll$. Taking into account Equation \eqref{pi-star} we conclude that the operators  $\Phi^t_{S,f,\la}$ and $\Phi^t_{S,f,\la'}$ correspond to, respectively, ${\cal T}_{x}$ and ${\cal T}_{x'}$ such that $x\neq x'$. Now \eqref{TT} gives us immediately
\[
||\Phi^t_{S,f,\la''}-\Phi^t_{S,f,\la'}||\geq 2.
\]      
In oder words, the net $\{\Phi^t_{S,f,\la}\}$ is not convergent in the uniform topology on $\cal B$. 

To prove Statement 2 note that if $\Phi^t_{S,f,\la}$ is the unit of $\cal B$ then the evaluation of the operator on any state gives $1$ as the result. On the other hand, there exists $\la_0$ such that for every $\la\geq\la_0$ the operator $\Phi^t_{S,f,\la}$ is not the unit and corresponds to a translation operator ${\cal T}_{x}$ on $L^2(\R^N,dx)\cong \h_\la$ such that $x\neq 0$. Then for $\rho=\{\rho_\la\}$ we have 
\[
\langle \Phi^t_{S,f,\la},\rho\rangle=\langle {\cal T}_{x},\tl{\rho}\rangle,
\]   
where $\tl{\rho}$ is a density matrix on $L^2(\R^N,dx)$ corresponding to $\rho_\la$. There exists an orthonormal basis $\{\psi_n\}$ of  $L^2(\R^N,dx)$ and real non-negative numbers $\{\tl{\rho}_n\}$ such that
\[
\tl{\rho}=\sum_n \tl{\rho}_n \ket{\psi_n}\bra{\psi_n} \ \ \text{and} \ \ \sum_n \tl{\rho}_n=1.
\]
Given $n$, the Schwarz inequality  gives us 
\[
|\scal{\psi_n}{{\cal T}_x\psi_n}|\leq ||\psi_n||\,||{\cal T}_x\psi_n||.
\]
Suppose now that both sides of the above formula are equal. This, however, can happen only if ${\cal T}_x \psi_n=\alpha\psi_n$. Since ${\cal T}_x$ is unitary $|\alpha|$ has to be equal to $1$, which means that $\overline{\psi}_n\psi_n$ is a periodic function with the period $x\neq 0$ and thereby $\psi_n$ cannot belong to $L^2(\R^N,dx)$. By virtue of this contradiction 
\[
|\scal{\psi_n}{{\cal T}_x\psi_n}|<||\psi_n||\,||{\cal T}_x\psi_n||=1
\]
and 
\[
|\langle \Phi^t_{S,f,\la},\rho\rangle|=|\langle {\cal T}_{x},\tl{\rho}\rangle|=|\sum_n \tl{\rho}_n\scal{\psi_n}{{\cal T}_x\psi_n}|\leq \sum_n\tl{\rho}_n|\scal{\psi_n}{{\cal T}_x\psi_n}|<1.
\]

Let finally justify Statement 3. Consider the state $\rho=\{\rho_\la\}$ occurring in Equation \eqref{lim-zero}. For $\la=\tL$, where $\tL$ is generated by the independent loops $\{l_1,\ldots,l_N\}$  we have
\begin{multline}
\langle \Phi^t_{S,f,\la},\rho\rangle=\langle\exp(\,it\,T_\la(\hat{\phi}_{S,f,\la})\, ),\rho_\la\rangle=\\=\langle\exp(\,it\,T_{\tL(l_1)}(\hat{\phi}_{S,f,{\tL(l_1)}})\, ),\rho_{\tL(l_1)}\rangle\ldots\langle\exp(\,it\,T_{\tL(l_N)}(\hat{\phi}_{S,f,{\tL(l_N)}})\, ),\rho_{\tL(l_N)}\rangle=\\=\langle \Phi^t_{S,f,\tL(l_1)},\rho\rangle\ldots\langle\Phi^t_{S,f,\tL(l_N)},\rho\rangle.
\label{T-rho}
\end{multline}
There exists an {\em uncountable} set $\{l\}$ of loops such that $(i)$ they generate pairwise distinct tame groups $\{\tL(l)\}$ and $(ii)$ for every loop $\phi_{S,f}h_l\neq 0$---the last property means that $\Phi^t_{S,f,\tL(l)}$ is not the unit of $\cal B$. Consequently, there exists a sequence $(\tL(l_n))$ of pairwise distinct groups such that for every $n$
\[
|\langle\Phi^t_{S,f,\tL(l_N)},\rho\rangle|\leq\delta<1.
\]
Let $\la_n=\tL_n$ be a tame group generated by the loops $\{l_1,\ldots,l_n\}$. Then it follows from \eqref{T-rho} that for every $\la\geq\la_n$ 
\[
|\langle\Phi^t_{S,f,\la},\rho\rangle|\leq\delta^n.
\]   
This implies \eqref{lim-zero} immediately.


\end{document}